\documentstyle[epsf,prb,aps,eqsecnum,tighten,preprint]{revtex}

\input epsf
\begin{document}
\draft

\title{Stripes, Vibrations and Superconductivity}
\author{A.~H.~Castro Neto$^*$}

\address{Department of Physics,
Boston University, Boston, MA, 02215}

\date{\today}
\maketitle

\begin{abstract}
We propose a model of a spatially modulated collective charge state (CCS) 
of superconducting cuprates 
\cite{first}. The regions of higher carrier density (stripes) 
are described in terms of one-dimensional (1D) interacting fermions and the 
regions of lower density as a two-dimensional (2D) interacting bosonic gas of 
$d_{x^2-y^2}$ hole pairs. The interactions among the elementary excitations are
repulsive and the transition to the superconducting state is driven by decay 
processes. Vibrations of the CCS and the lattice, although not participating 
directly in the binding mechanism, are fundamental for superconductivity. 
The superfluid density and the lattice have a strong tendency to modulation 
with wave-vectors $(\pi/a,0)$ and $(0,\pi/a)$ implying 
a still unobserved dimerized stripe phase in cuprates. The phase diagram 
of the model has a crossover from 1D to 2D behavior and
a pseudogap region where the amplitude of the order parameters are finite 
but phase coherence is not established. We discuss the nature of the 
spin fluctuations and the unusual isotope effect within the model.
\end{abstract}

\bigskip


\newpage

\narrowtext

\section{Introduction}
\label{intro}

The experimental evidence for charge and spin inhomogeneities in 
superconducting
cuprates has been accumulating in the last few years 
\cite{experiments,millis}. 
It is believed that
these inhomogeneities are probably the result of strong competing 
electron-electron and electron-lattice interactions. The strong interactions
can be hinted from the fact that cuprates are related to Mott insulators, not to metals.
From the theoretical point of view the major problem has been the description 
of the doped Mott insulator. It is this state that
finally evolves into a superconducting state. 
The origin of the inhomogeneities can be diverse. 
Disorder and the natural charge localization effects that occur
close to an insulating Mott state are natural possibilities.
Furthermore, there is an emerging
point of view that inhomogeneities may be in the heart of the superconducting 
phase. The difficulty in describing the superconducting state 
may be related with the fact that an important part of the physics occurs in
real space. This should be contrasted with a Fermi liquid description that
is dominated by the Fermi surface in momentum space. If the inhomogeneities
observed experimentally are essential for the description of the superconducting
state then a dual description in terms of real and momentum space is certainly
required.
 
The search for a general principle or a special (maybe hidden) 
symmetry that facilitates
the understanding of these materials is always desirable. It turns out that
the intrinsic complexities of these systems are enormous. 
In fact, the experimental evidence seems to be that all the degrees of freedom,
charge, spin, lattice, participate in a
fundamental way in their physics. Unfortunately, we may have to sacrifice the 
simplicity of description that helped us to understand the basic physics
of many systems in the past.
Besides the strong local interactions that induce large
charge fluctuations, 
long-range interactions associated with the insulating state
can suppress the same degrees of freedom. We believe that it is the
interplay between these two forces that is fundamental for the understanding
of the physics in these systems. 
In this paper we assume that the physics 
in cuprates can be divided at least into two length scales. 
On the one hand there is the short length (high energy) scale that
is associated with the behavior of the electrons at atomic distances. 
This piece of physics
is dominated by the strong local interactions that lead to the
Mott insulator and it is well described by models of strongly
correlated electrons such as the Hubbard or the t-J model. 
This is a ``highly'' quantum regime where quantum
fluctuations at atomic scales determine the physical response.
Since the short length scale is usually smaller
than the size of systems studied numerically, these studies can provide a
good physical insight. Various
analytical approaches to these strongly coupled models exist in 2D
\cite{ash,dicastro,bob} and although they can also provide insight into the 
main physical aspects of the problem 
they are unreliable in dealing with quantum fluctuations. 
In order to
acquire intuition one is forced to look at purely academic models that
can be solved beyond mean-field. 
This is the case of the various studies of
the Ising $t-J$ model where quantum spin fluctuations are suppressed by a 
large anisotropy
in spin space \cite{tjzmodel}, the spin-fermion models where the spins
are treated classically \cite{moreo} or models with large number of
components \cite{vojta}.
On the other hand, the long wavelength physics of
the problem is dominated by long-range interactions such as the Coulomb, 
Casimir \cite{casimir} or
entropic interactions that are essentially classical in origin 
\cite{nils,zaanen}. 
Analytical treatments
of these interactions have always been difficult even in 
statistical mechanical
problems in 1D. Therefore, from the theoretical point of view there are
technical complications at all length scales.  
Any theoretical treatment that intends to start 
from the microscopic picture from the beginning will encounter tremendous 
difficulties.

In this paper we follow a semi-phenomenological approach. 
At the short length scales 
we have known for a long time that doped Mott insulators 
have a big ``aversion'' to
charge homogeneity \cite{mott}. This happens because there is a large 
loss of magnetic energy
when charge delocalizes. Even before superconductivity was discovered 
in cuprate oxides the concept of {\it strings} and quantum confinement of
holes was well understood: 
when a hole moves in an antiferromagnet it produces a string
of overturned spins with an energy growing with its length \cite{strings}.
When the hopping energy, $t$, is smaller than the antiferromagnetic exchange, $J$, strings are the dominant processes in antiferromagnets 
(although higher energy delocalizing
processes called Trugman loops are also possible \cite{trugman}). 
When two holes are injected into an
antiferromagnet there is a gain in magnetic and kinetic energy if they 
move together instead of separately. Strings are suppressed 
because the second hole heals the string generated by the first one.
This picture of the diluted Mott insulator is generic. It can be obtained
analytically in the Ising $t-J$ model and is also 
observed in numerical simulations of the SU(2) models as well \cite{tjzmodel}. 
Because of the tendency to segregate charge, Mott insulators are at the edge
for phase separation into 
hole-rich and hole-poor regions \cite{phaseparation}. Still
to this date there is a heated debate about the 
border to phase separation in the phase diagram of the $t-J$ model 
\cite{numericaldebate}. More recently the concept of {\it stripes}
in the $t-J$ model 
has emerged due to density matrix renormalization group (DMRG)
calculations by White and Scalapino (WS) \cite{whitescalapino}. 
These simulations show clear signs 
of charge order. In the WS picture when holes are doped into the 
antiferromagnet
they first form pairs that condense into lines of charge with
magnetic anti-phase domain walls (ADW). This condensation process, 
as shown in analytical approaches
to the Ising $t-J$ models \cite{sasha}
and in other numerical studies \cite{elbio}, 
has to do with the gain in kinetic energy of a {\it single} 
hole due to the formation
of an ADW (moreover, Trugman loops are supressed in this case \cite{sasha}). 
That is, for a finite linear density of holes (vanishing
2D density) the gain in kinetic energy by making a
domain wall is enough to compensate for the loss of magnetic energy due to the
formation of a magnetic defects. We should stress that we are not 
concerned with the problem of the phase diagram of the 
$t-J$ model because this 
phase diagram most certainly {\it cannot} be the phase diagram 
of cuprate oxides. There are 
many important interactions in cuprates that are not included 
in the $t-J$ model.
We will argue below that coupling to the lattice, for instance, 
is relevant for 
the experimental phase diagram. We assume 
that the tendency of the diluted Mott insulator 
to form bound state pairs of holes is universal. 

Long-range interactions and their eventual screening are probably 
the key to understand 
the phase diagram of cuprates \cite{branko}. At low doping, because of 
the existence of a
large charge gap in the Mott insulator, charge dynamics is suppressed and
as a consequence dynamical screening as well. All the physics
rests in the spin degrees of freedom. Thus, when a small concentration of hole
is introduced 
into the CuO$_2$ layers long-range interactions should play a major role. 
These long-range interactions will be effective in suppressing the tendency
to phase separation as stressed by Emery and Kivelson (EK) and
collaborators \cite{emerykivelson,erica,granath} 
and will lead to the generation of a finite length scale, $L_D$,
associated with domain size \cite{brep}. In the 
absence of a lattice the long-range forces produce 
blobs or lakes of charge with characteristic size $L_D$. It turns out,
however, that in transition metal oxides the coupling to the lattice is strong
and therefore the symmetry of rotations in real space is broken. 
Charge modulation is therefore the final result of the tendency to phase
separation and strong lattice coupling. The formation of
this charge modulated phase is {\it not} the result of a Fermi surface 
singularity as
in the case of 1D charged density waves (CDW). 
In fact, the CCS state described here has strong similarities with the
metallic
2D charge density wave states observed in dichalcogenides \cite{cdw} 
that have very little resemblance with the 1D CDW states that are usually
insulating \cite{gruner}. 
This state is probably better described as
coming from the quantum melting of an anisotropic Wigner crystal 
(or Bragg glass) \cite{branko} or an exotic CDW state \cite{chetan}. 
In classical
statistical mechanics there are many analogues called liquid crystal phases:
nematic, smectic, hexatic, etc \cite{lubensky}. 
These are phases with long-range orientational order
in the absence of translational order. 
Another more mundane phase where translational
symmetry is broken in 2D is an anisotropic {\it membrane} phase where 
translational
order is broken along the principal vectors of the lattice \cite{nils}.

Let us consider a static charge modulated phase induced along the $y$ direction
in the 2D system with periodicity $N a$ where $a$ is the lattice spacing.
The charge density, $\rho({\bf r})$, 
can be written as (we use units such that $\hbar=k_B=1$)
\begin{eqnarray}
\rho({\bf r}) = \rho_0 + \langle \Phi({\bf r}) 
\rangle \exp\left\{i \frac{2 \pi}{N a} y
\right\} + c.c.
\label{rhor}
\end{eqnarray}
where $\rho_0$ is the background density, and 
$\Phi$ is the complex order parameter that can be rewritten
in terms of an amplitude $|\Phi|$ and phase $u$:
\begin{eqnarray}
\langle \Phi({\bf r})\rangle  = |\Phi| 
\exp\left\{-i \frac{2 \pi}{N a} u({\bf r}) \right\} \, .
\end{eqnarray}
The lines of constant phase at wave number
$2 \pi/(N a)$ are describe by
\begin{eqnarray}
\phi = \frac{2 \pi}{N a} (y - u) = 2 \pi n
\label{cp}
\end{eqnarray}
where $n=0,\pm 1,\pm 2,...$. In a inert background (such as a Fermi liquid) 
the local modulations of the density can lead to gaps at the Fermi surface
but the system as a whole remains homogeneous. In
a Mott insulator this is not necessarily so. 
In the regions of low charge density the 
Mott insulator is essentially untouched. In these regions it is energetically
advantageous 
for the system to form bound states of holes with
d$_{x^2-y^2}$ symmetry (exactly like in a finite cluster or ladder)
with a gain magnetic energy \cite{elbio,dwave,andrey}. 
In the regions where charge density is 
large it is energetically more favorable for the system to have a 
gain of kinetic energy \cite{sasha} that liberates the holes to move 
as single particles 
and create ADW. Thus, the formation of a charge
ordered state has to be accompanied by a change in the spin structure 
of the system with the creation of incommensurate spin fluctuations.
The mechanism of gain of energy (kinetic or magnetic) depends strongly on the
amount of charge density. Moreover, deformations of amplitude of the order 
parameter of
the CCS are always energetically costly. Therefore, phase fluctuations
are the low energy excitations in such systems. 
For a classical smectic phase, for instance,
the free energy is given by
\begin{eqnarray}
F_S &=& \frac{1}{2} \int d^2 r \left[B (\partial_{||} u_S)^2+ K 
(\partial^2_{\perp} u_S)^2\right]
\nonumber
\\
&=& \frac{1}{2} \sum_{{\bf k}} \omega_{{\bf k}} |u_{S,{\bf k}}|^2 
\end{eqnarray}
where $B$ is transverse stiffness of the smectic and $K$ is the splay elastic
constant. Here $||$ and $\perp$ indicate the gradient of the function in the
directions parallel and perpendicular to the ordering direction, respectively.
Moreover,
$\omega_{S,{\bf k}} = B k_{||}^2 + K k_{\perp}^4$ is the classical smectic
dispersion relation. A membrane phase, on the other hand, would be
described by $\omega_{M,{\bf k}} = \sqrt{B^2_{||} k_{||}^2 + B_{\perp}^2 k_{\perp}^2}$
where $B_{||}$ ($B_{\perp}$) are the longitudinal (transverse) compressibilities.
In fact, because the charge modulated state is composed of electrons
we expect this modulated phase to be quantum in nature, that is, a {\it quantum liquid
crystal} \cite{qualiqcrys}.

Notice that such a smooth charge distribution as proposed in (\ref{rhor})
is different from the
stripe phases discussed in the context of insulating cuprates and nickelates. 
In insulating nickelates static
charge order is observed in neutron scattering and many other 
experiments \cite{nickelates}. 
The key point is that higher {\it harmonics} of the fundamental Bragg peaks
associated with static order are observed in these systems. 
It implies sharp, well defined and isolated domain wall
structures in real space. In superconducting systems where inelastic neutron
scattering peaks are observed at equivalent positions there are {\it no}
harmonics observed, even when the fundamental peaks are rather sharp. 
In La$_{1.48}$Nd$_{0.4}$Sr$_{0.12}$CuO$_4$ (LNSCO) where Tranquada and collaborators observed
quasi-static peaks in neutron scattering 
there are no signs of higher harmonics \cite{tranquada}.
This experimental fact signals to a smooth variation of the charge-spin densities in the system.
Thus, the idea of well-defined, non-interacting stripes is misleading in the context
of superconductivity. Instead one should think of the ``stripes'' as a complex collective
state that is driven by competing local interactions and cooperative 
long-range forces.

The importance of the lattice degrees of freedom has been experimentally verified in
essentially all cuprate superconductors and their insulating 
relatives \cite{experphonons,egami}.
At this point in time most of the theoretical approaches either focus 
on electron-electron
interactions and overlook the importance of lattice degrees of freedom 
or mainly electron-phonon interactions \cite{alexandrov} 
disregarding the importance of the
strong (short- and long-range) interactions in the problem. Some 
mean-field approaches, however, have stressed the 
importance of electron-lattice interaction in the context of 
nickelates \cite{dicastro,alan}. 
We believe that electron-electron and electron-lattice interactions
are equally important because of charge neutrality.
Charge neutrality implies that a charge modulated state such
as the one defined in (\ref{rhor}) has to be strongly coupled to the lattice.
One would expect under general considerations that fluctuations of
the CCS to appear in the phonon spectrum. Indeed,
consider the classical elastic lattice free energy:
\begin{eqnarray}
F_P = \frac{1}{2} \sum_{{\bf k}} \omega_{P,{\bf k}} |u_{P,{\bf k}}|^2
\end{eqnarray}
where $u_{P,{\bf k}}$ is the lattice deviation from equilibrium position and
$\omega_{P,{\bf k}}$ the phonon dispersion relation. In the 
leading order the coupling between the smectic 
and lattice is quadratic in the displacements and
can be written as
\begin{eqnarray}
F_C = \sum_{{\bf k}} C_{{\bf k}} u_{{\bf k}} u_{P,-{\bf k}} + c.c.
\end{eqnarray}
where $C_{{\bf k}}$ are the coupling constants. 
The problem of the CCS plus
phonons can be solved exactly by a simple linear combination of the 
displacement fields. There are two branches of 
excitations with frequency
given by
\begin{eqnarray}
\Omega_{\pm,{\bf k}} = \frac{1}{2} \left[\omega_{S,{\bf k}} + \omega_{P,{\bf k}}
\pm \sqrt{(\omega_{S,{\bf k}} - \omega_{P,{\bf k}})^2+4 |C_{{\bf k}}|^2}\right]
\end{eqnarray}
leading to a splitting of the vibrational modes.
Therefore, the phonon spectrum should be directly affected by the presence of
a CCS. Thus,
in dealing with the fluctuations of the CCS we have to consider
the renormalizations of such fluctuations by the lattice. The full quantum
mechanical problem can be quantized exactly like
phonons in ordinary crystalline solids and the vibrations of the CCS can be described by 
\begin{eqnarray}
H_V = \sum_{{\bf k},\alpha} \Omega_{\alpha,{\bf k}} b^{\dag}_{\alpha,{\bf k}} b_{\alpha,{\bf k}} \, .
\end{eqnarray}
where  $b_{\alpha,{\bf p}}$ ($b^{\dag}_{{\bf k},\alpha}$) is the annihilation (creation) operator 
for quantum vibrations of the CCS with momentum ${\bf k}$ in
the branch $\alpha$
and energy $\Omega_{\alpha,{\bf k}}$.

Although the description of the collective state 
is rather simple because of its
gaussian nature, the description of its internal degrees of freedom 
(associated
with the short length scales) is more complex. 
In what follows we will make an {\it ad hoc} assumption that the
internal charge excitations of this collective state can be divided into two 
main
groups. Firstly, in the regions where the charge density is large 
(that is, given
by the lines of constant phase in (\ref{cp})) ADW exist
due to the local gain in kinetic energy. These regions we call stripes. 
These high density regions are characterized by single particle
excitations (not bound states!) and they are essentially confined to 
1D lines because of the potential induced by strings \cite{sasha}. 
This highly anisotropic electronic
fluid should be interacting because of the phase space constraints 
and in the absence of tunneling between stripes it is described by a
Luttinger liquid \cite{luttinger}. 
In the regions of low density 
(in the middle of the antiferromagnetic ladders) single
particle excitations are suppressed and bound states of holes are energetically
more favorable. The simplest
of them is a bound state of a pair of holes with d$_{x^2-y^2}$ 
symmetry \cite{dwave}. 
Larger bound state structures like quartets are unlikely
to contribute because their quantum dynamics is exponentially suppressed 
(tunneling matrix elements decay exponentially with the number of particles). 
Moreover, we assume that this gas of bosons
is essentially isotropic and only weakly coupled to the stripe fermions. The main
reason of the weak coupling is related with the ``string healing'' process that
generates the pairs in first place, that is, the bosons are essentially insensitive
to the magnetic structure including the ADW. 
We argue, however, that it is the weak coupling between bosons and
stripes that ultimately produces superconductivity in the system. 

\begin{figure}
\epsfysize6 cm
\hspace{0cm}
\epsfbox{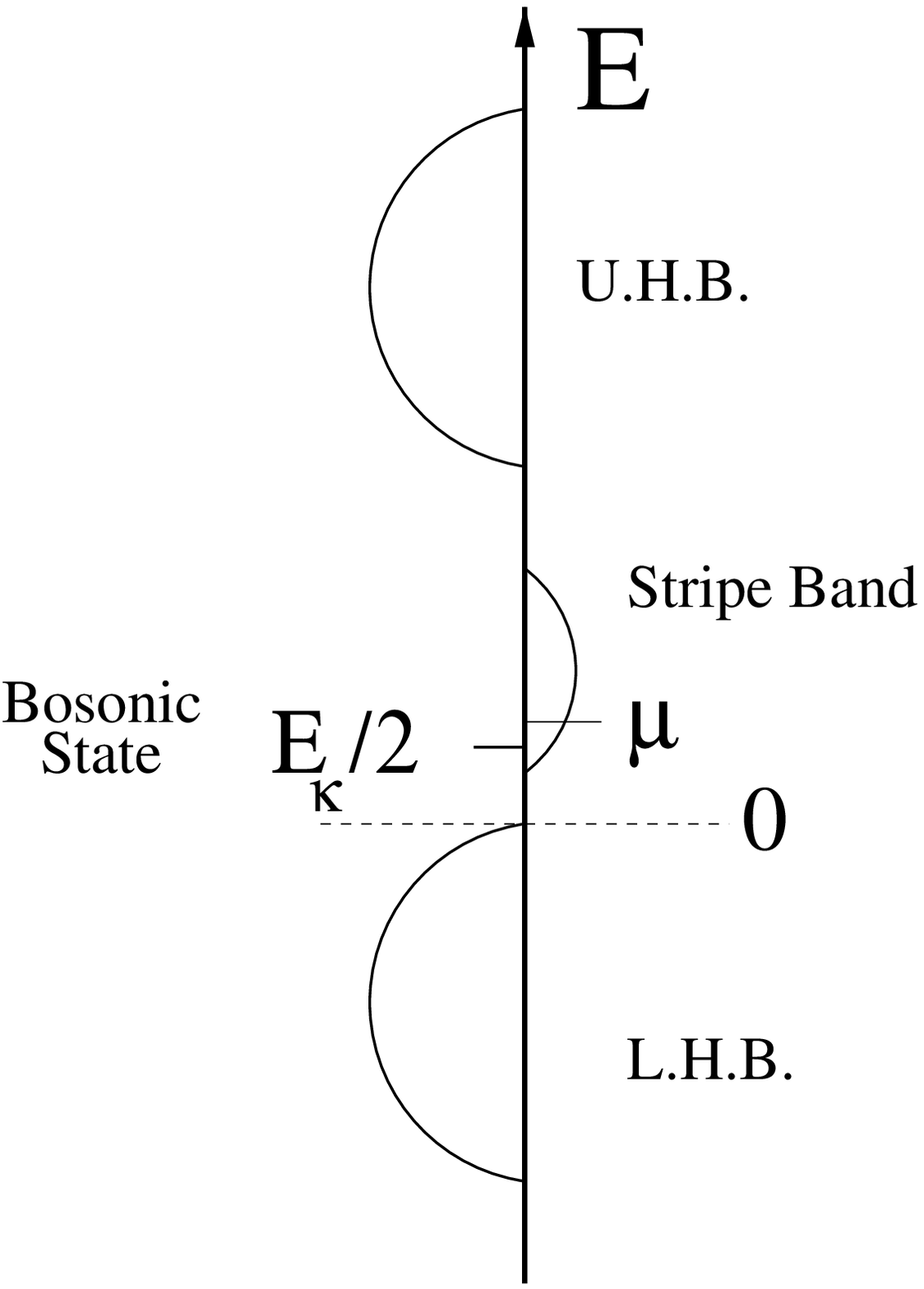}
\caption{Schematic plot of the density of states as a function of
energy.}
\label{dos}
\end{figure}

From the electronic point of view the situation is illustrated in Fig.\ref{dos}.
The Mott insulating state can be described by a filled lower Hubbard band (L.H.B.)
and an empty upper Hubbard band (U.H.B.) that are separated by a large energy scale.
In between these two bands there is a single electron band associated with
the stripe fermions that we call the stripe band. The bound states of holes 
exist due to 
the transfer of spectral weight from the L.H.B. to a level above it 
with energy $E_{{\bf k}}/2$ (where ${\bf k}$ is
the momentum of the boson as we discuss below). The binding energy
of the holes is the energy difference between the boson level and the top of the
L.H.B.. The binding energy can be seen as mediated by the exchange of 
paramagnons \cite{pines} and therefore is of the order of the 
characteristic magnetic energy in the
problem \cite{numericpair,sushkov}. In the undoped system the characteristic
energy is simply the exchange constant $J$ but as doping increases 
the magnetic energy scales are reduced driving the system from the Mott
insulator to a more ordinary Fermi liquid state.

\begin{figure}
\epsfysize6 cm
\hspace{0cm}
\epsfbox{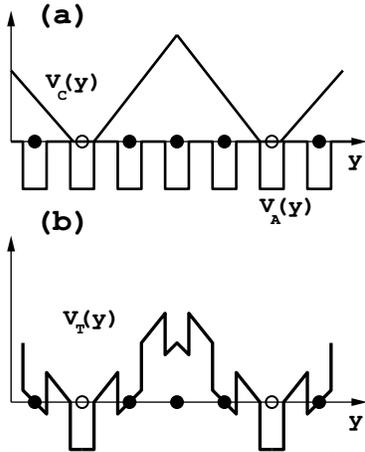}
\caption{Effective potential felt by the carriers: (a) $V_c(y)$ is
the string potential generated by the antiferromagnetic background and
$V_A(y)$ is the atomic potential generated by the lattice; (b) resultant
potential felt by the carriers.}
\label{potential}
\end{figure}

The measure of the magnetic energy is the spin stiffness,
$\rho_s$, of the magnetic background. A finite spin stiffness produces the
confining potential for the charge carriers of the form \cite{sasha}:
\begin{eqnarray}
V_C(y) = \frac{\rho_s}{a} |y|
\label{confine}
\end{eqnarray}
(modulus $N a$) as shown in Fig.\ref{potential}(a) ($V_C(y+N a) = V_C(y)$). 
The total potential as seen by the holes is a superposition of the atomic
potential of the lattice, $V_A(y)$ ($V_A(y+a)=V_A(y)$),  
and the magnetic confining potential of the strings.
In Fig.\ref{potential}(b) we show in a simple Kronig-Penney picture
the result of the superposition of these two potentials.
When the temperature is larger than $\rho_s$ the holes are essentially
deconfined. 
Since the hole-pairs only exist at
temperatures below $\rho_s$ it is important to estimate this
energy scale. In order
to do so let us consider the situation in Fig.\ref{anisotropy}(a) 
where the exchange within the
antiferromagnetic regions is $J$ but across the stripe it is $J'<J$. 
At long wavelengths
the problem maps into a spatially anisotropic Heisenberg model with
exchanges $J_x$ and $J_y$ as shown in Fig.\ref{anisotropy}(b) \cite{hone}. 
While $J'/J$ is a short
length scale property and depends on the microscopic details \cite{sasha} 
the ratio
\begin{eqnarray}
\alpha = \frac{J_x}{J_y}
\label{alpha}
\end{eqnarray}
depends on the long wavelength properties and determines the region 
of stability for
antiferromagnetic order \cite{hone}. Moreover, the effective
spin stiffness of the magnetic background depends directly on $\alpha$. 
Assuming that
$J_y = J$ in Fig.\ref{anisotropy}(b) 
(so that the exchange along the stripe is not
modified by the presence of the stripes) it is easy to show that \cite{hone}
\begin{eqnarray}
\rho_s(\alpha) = J S^2 \sqrt{\alpha} \, .
\label{rhosa}
\end{eqnarray}
This choice for the exchange constants was used in ref.\cite{duin} and
seems to explain well the data in LNSCO \cite{tranquadamag}. 
In the low hole doping regime ($x<0.02$) another choice has to be 
made since the magnetism is isotropic 
\cite{hone}. 
It is clear from (\ref{rhosa}) that a decrease in $\alpha$ reduces the
spin stiffness and can 
lead to a quantum phase transition to a spin liquid state \cite{wang}.
The main problem is how to relate $\alpha$ of the effective model 
of Fig.\ref{anisotropy}(b) to
the microscopic model of Fig.\ref{anisotropy}(a). We use a simple
scheme and compare the classical ground state energies 
of the two problems. In the case of
Fig.\ref{anisotropy}(a) the Hamiltonian is simply
\begin{eqnarray}
H = \sum_{i,j} J_{i,j} {\bf S}_i \cdot {\bf S}_j
\end{eqnarray}
where $J_{i,j}= J$ in the antiferromagnetic ladders and $J_{i,j}=J'$ across the ADW.
Let us rewrite $J' = \gamma J$ where $0<\gamma<1$ is a microscopic
quantity that depends on details of the problem. 
With this parameterization it is clear that the classical
ground state energy can be written as
\begin{eqnarray}
E_0 = E_{AF}(J) + (1-\gamma) J S^2 N_{stripes} N_s
\label{e0a}
\end{eqnarray} 
where 
\begin{eqnarray}
E_{AF}(J) = - 2 J S^2 N_{spins}
\end{eqnarray}
is the classical energy of an isotropic antiferromagnet with $N_{spins}$ 
spins and
exchange $J$. $N_s$ is the
number of sites in each direction ($N_s^2$ is the total number sites),
$N_{stripes}=N_s/N$ is the number of stripes in the system ($N$ is the
separation between stripes in lattice units), and
$N_{spins} = N_s^2-N_s \times N_{stripes}$ is the number of
spins not residing in stripes ($N_{spins} = N_s^2(1-1/N)$). Thus, from
(\ref{e0a}) one gets,
\begin{eqnarray}
\frac{E_0}{N_s^2 J S^2} = -2 + \frac{3-\gamma}{N} \, .
\label{e0afinal}
\end{eqnarray}
On the other hand, for the effective model shown on Fig.\ref{anisotropy}(b) with the
same number of sites we would have
\begin{eqnarray}
E_0 = - (J_x+J_y) S^2 N_s^2 = -J (1+\alpha) S^2 N_s^2
\label{e0b}
\end{eqnarray}
where we used (\ref{alpha}). Comparing (\ref{e0afinal}) and (\ref{e0b}) we find
\begin{eqnarray}
\alpha = 1 - \frac{3-\gamma}{N} \, .
\end{eqnarray}
This result implies from (\ref{rhosa}) that
\begin{eqnarray}
\rho_s(N) = J S^2 \sqrt{1 - \frac{3-\gamma}{N}} \, .
\label{rhosn}
\end{eqnarray}
Them the spin stiffness vanishes
at a critical distance $N_m$ between stripes given by 
\begin{eqnarray}
N_m = 3 -\gamma .
\label{nm}
\end{eqnarray} 
Since $0<\gamma<1$ we see that $2 < N_m < 3$ (quantum fluctuations
increase the value of $N_m$ \cite{hone}).
The binding energy of the hole-pairs is
proportional to $\rho_s$ and no bosons can exist at zero 
temperature when
$N<N_m$. Moreover, we identify a temperature scale 
$T^*(N) \approx \rho_s(N)$ above
which no bosons exist ($T^*(N_m)=0$). 
So we relate the formation of the hole pairs as 
the ``pseudogap'' energy scale
observed in many different experiments \cite{loram}. Our theory, however,
breaks down when $T > T^*$.

In the non-interacting picture (with no hybridization
between bosons and fermions) we can have two possibilities.
In Fig.\ref{dos} the boson state is empty of holes (filled with electrons)
and the stripes are partially filled up to the chemical potential energy $\mu$.
Because bosons and fermions are in thermodynamic equilibrium we work in
the grand-canonical ensemble and keep the chemical potential fixed by 
letting the number
of particles fluctuate. 
Consider the case when 
\begin{eqnarray}
\mu > E_{{\bf k}}/2 \, .
\nonumber
\end{eqnarray}
The boson state is unoccupied and there are no
bosons in the system. 
As the chemical potential is reduced (more
holes are introduced) the stripes empty and further reduction of the chemical
potential pins the energy at the boson level since as more holes are added 
they can only produce bound states. Thus, there is a continuous transfer of spectral weight
from the L.H.B. to the boson level. 
On the other hand, if the binding energy of the holes
increases nothing happens until $\mu = E_{{\bf k}}/2$. 
At this point there would be
bosons and stripes co-existing with each other. When
the binding energy is increased further 
the chemical potential follows since bosons become
converted into stripe fermions. Once again the chemical potential is 
pinned at the
boson level. Finally when the boson level reaches the top of the stripe band 
the stripes
are completely filled and the ADW disappears. 
In fact, because of the
loss of kinetic energy the domain wall disappears even before this
limit is reached \cite{sasha}. 
Therefore, we assume that
\begin{eqnarray}
2 \mu - E_{{\bf k}} \geq 0 \, .
\label{condition}
\end{eqnarray} 
Moreover, it is clear from this picture that the
spin degrees of freedom that are responsible for the large magnetic response in
these systems leave on the L.H.B. and are separated in energy from the
charge degrees of freedom. Therefore, the magnetism can be effectively ``traced out''
of the problem since it will only lead to kinematic renormalizations of the various
parameters (in other words, the spins follow the charge).

The mechanism for superconductivity discussed in this work requires the coupling of
stripe fermions via the exchange of bosons. One can think of this mechanism as the exchange
of stripe ``pieces''. Coherence between the fermions requires the 
exchange of real bosons. 
The simplest mechanism for
exchange is the decay of the bosons (since they are composite particles) into
fermionic degrees of freedom at the stripes. This kind of mechanism can
occur when two systems with very different ground states are separated by
an interface. In fact, it was proposed long ago that a mechanism of this
sort could generate superconductivity at a metal-semiconductor interface
(also called exciton superconductivity)\cite{bardeen}. Although this
kind of proposal has generated controversy in the past 
\cite{controversy} there are good indications that they may be good
candidates in the case of cuprates \cite{geballe}.
Moreover, the process described here is similar to
the ``proximity effect'' mechanism proposed by EK if we 
neglect retardation effects
due to the boson motion \cite{emerykivelson,erica,granath}. 
One of our main results is: if 
the stripes static, that is, if we disregard the fluctuations of the CCS, 
this process is suppressed! The ``proximity effect''
mechanism cannot happen with static stripes and real bosons. 
Therefore for coherence to be attained we have to introduce
fluctuations of the CCS.
This makes our model radically different from the proximity effect proposed by EK
since it involves distortions of the CCS and therefore of the lattice
as well. Moreover, unlike BCS, the phonons are not part of 
the binding mechanism that is driven by 
the exchange of composite fermion pairs. Vibrations, however, are fundamental for the
superconductivity.
A natural consequence of the mechanism is that fluctuations of
the order parameter are necessarily coupled to the fluctuations of the CCS 
and therefore vortices are coupled to dislocations of the stripe array. Because
the superfluid density is low, the interactions between 
topological defects ultimately determines the phase diagram.
Moreover, we claim that static stripes (as the ones observed obtained in Hartree-Fock
solutions of the $t-J$ and Hubbard models \cite{hartfock,ortiz} or 
DMRG \cite{whitescalapino}) 
should be insulating due to a 1D CDW 
instability along the stripe direction. 
Diagonal stripes do not couple to the bosons because they
are oriented along the nodes of the boson wave-functions.

Among other things we explain why phonon anomalies that have been observed in
neutron scattering \cite{experphonons} occur exactly at the same position in the Brillouin zone
where angle resolved photoemission (ARPES) \cite{arpes} observes the opening of the pseudogap
(that is, at $(\pi/a,0)$ and $(0,\pi/a)$). 
Moreover, we show that the lattice
distortions associated with the phonon anomalies are associated with the
modulation of the
superfluid density {\it perpendicular} to the stripes. 
The theory also predicts an isotope effect and the 
co-existence
of commensurate and incommensurate spin fluctuations. 
The critical temperature for superconducting order, $T_c$, is determined
by the interplay between topological defects associated with the superfluid
bosons (vortices and anti-vortices) and distortions of the CCS (dislocation
loops). The transition at finite temperatures is in the 3D XY universality
class.
We show that the amplitude of the order parameters is finite at
temperatures above $T_c$ up to $T^*$ in the pseudogap
region. Moreover, at $T=0$ there is a quantum phase transition as
a function of the separation between stripes, $N$, that is
directly related to the hole concentration $x$ ($N \approx 1/(2x)$). 
While the order parameters become finite for $N < N_{sp}$ long-range
order is only attained at $N=N_c<N_{sp}$. This quantum phase transition is
in the 2D XY in a magnetic field universality class. Thus, there is
a crossover region at $T=0$, $N_{c}<N<N_{sp}$, where topological
defects prevent long-range order to develop.

The paper is organized as follows: in the next section we introduce the bosonic
bound state of holes; in section \ref{stfe} we present the Luttinger liquid
representation of the stripe fermions; in section \ref{coupl} we discuss
the nature of the coupling between bosons and stripe fermions and argue that
vibrational degrees of freedom should enter explicitly; 
in section \ref{mfield} we solve the mean-field 
equations for the problem and determine the mean-field phase diagram; 
in section
\ref{phflu} we discuss the phase fluctuations of the
superfluid-superconducting state and how they relate to the phase fluctuations
of the CCS; in section \ref{hight} the nature of the high
temperature transition is discussed; section \ref{lowt} contains the results for the zero temperature
transition; section \ref{conclu} contains our conclusions.

\section{Antiferromagnetic Bosons}
\label{afb}

Consider the problem of two holes in an antiferromagnet. The energy 
is minimized by the formation a bound state that moves freely through
the system \cite{dwave,sushkov}. One can define an operator 
that creates such a state at the top
of the antiferromagnetic vacuum that is given by
\begin{eqnarray}
P^{\dag}_{{\bf k}} = \frac{1}{\sqrt{S}} \sum_{\bf q} \Psi_{{\bf k}}({\bf q}) 
c_{\uparrow,{\bf k}/2-{\bf q}} c_{\downarrow,{\bf k}/2+{\bf q}}
\label{pk}
\end{eqnarray}
where $c_{\sigma,{\bf k}}$ destroys an electron (creates a hole) 
with momentum ${\bf k}$ and
spin projection $\sigma$ ($\uparrow$ or $\downarrow$) in the antiferromagnet.
$S=(N_s a)^2$
is the area of the system. 
The wavefunction of the pair, $\Psi_{{\bf k}}({\bf q})$, is normalized 
($\sum_{q}|\Psi_{{\bf k}}({\bf q})|^2/S =1$) and in principle depends
on the relative and total momentum of the pair. This wave-function can be obtained variationally \cite{hamer} 
or by solving the Bethe-Salpeter equation for
the binding of two holes \cite{sashagain}. The dependence of the 
wave-function on the total momentum
is due to the fact that on the lattice the symmetry of the 
bound state varies with its center of mass
momentum and
depends strongly on the microscopic details.
For the SU(2) $t-J$ model it is known that the d$_{x^2-y^2}$-state has the 
lowest
energy \cite{dwave}
while in the Ising $t-J$ model the p-wave is the ground state 
\cite{sushkov,ladder}. 
The matter of the fact is that the s-wave state is always
the state with highest energy and the reason is fairly simple to
understand: the strong
local repulsion requires the wave-function of the pair to be centered 
at different sites. In other words, s-wave bound states are
suppressed by the magnetism. 
Here we assume the pairs to have d$_{x^2-y^2}$ symmetry.
 
It is easy to see that operator defined in (\ref{pk}) is not completely bosonic
because of its composite nature \cite{lipkin}. 
Since the electrons obey anti-commutation relations 
($\{c_{\sigma,{\bf q}},c^{\dag}_{\sigma',{\bf q'}}\} = 
\delta_{{\bf q},{\bf q'}} \delta_{\sigma,\sigma'}$) 
it is easy to show that commutation relations of the bosons is given 
by:
\begin{eqnarray}
\left[P_{{\bf k}},P^{\dag}_{{\bf k'}}\right]&=&\delta_{{\bf k},{\bf k'}}
- D_{{\bf k},{\bf k'}}
\label{bos}
\end{eqnarray}
where
\begin{eqnarray}
D_{{\bf k},{\bf k}} &=& \frac{1}{S} 
\sum_{{\bf q}} \left(|\Psi({\bf k}/2-{\bf q})|^2 n_{h,\uparrow,{\bf q}}+
|\Psi({\bf k}/2+{\bf q})|^2 n_{h,\downarrow,{\bf q}}\right)\, .
\label{dkk}
\end{eqnarray}
Here $n_{h,\sigma}$ is the number operator for holes ($1-n_{c,\sigma}$ where 
$n_{c,\sigma}$ is the number
operator for electrons).
Since the magnitude of $D_{{\bf k},{\bf k}}$ is proportional to density of holes in the antiferromagnet,
$D_{{\bf k},{\bf k}}$ is smaller than $x$, the total number of holes in the
system. Moreover, the largest fraction of holes is actually residing in the stripes. 
Since (\ref{dkk}) is much smaller than one, the violation of the bosonic commutation
relations can be disregarded and we can treat the bound states as real bosons.
At this point it is convenient to count the number, $N_e$, of fermions in the system.
Assuming that there are $N_s/N$ stripes of size $N_s$ (in lattice units) the
number of antiferromagnetic sites is $(N_s-N_s/N) \times N_s$ and 
the total number of fermions is: 
\begin{eqnarray}
N_e = N_s^2(1-1/N) - 2 \sum_{{\bf k}} P^{\dag}({\bf k}) P({\bf k})
+ \sum_{{\bf k},\sigma}\psi^{\dag}_{{\bf k},\sigma} \psi_{{\bf k},\sigma}
\label{totn}
\end{eqnarray}
where the factor of $2$ comes from the composite nature 
of the bosons and the last term is the number of fermions in the stripes
(see below).

The Hamiltonian of the composite bosons can be written as
\begin{eqnarray}
H_P = \sum_{{\bf k}} (-E_{{\bf k}}+2 \mu) P^{\dag}_{{\bf k}} P_{{\bf k}} + 
\frac{U}{2} \sum_i N_i^2
\label{hp}
\end{eqnarray}
where $E_{{\bf k}}$ is the dispersion relation of the bosons 
and $U$ is the local repulsion between pairs. 
In (\ref{hp}) $\mu$ is the chemical potential for the
electrons and it comes with a negative sign because the chemical energy appears
as $-\mu N$ with the total number of electrons given in (\ref{totn}). 

Suppose there are no other interactions in the problem (that is, stripe fermions
and d$_{x^2-y^2}$ bosons do not interact). The Hamiltonian (\ref{hp})
is the so-called Bose-Hubbard model. This problem
has been studied to a great level of detail 
and its phase diagram is well-known \cite{matthew}.
Let us consider the situation close to the minimum of $E_{{\bf k}}$ at, say,
${\bf k}={\bf K}$. Expanding close to this point one can rewrite the 
Hamiltonian for the bosons as
\begin{eqnarray}
H = \sum_i \left[(-E_{{\bf K}}+2 \mu) N_i + \frac{U}{2} N_i^2\right] 
- t_B \sum_{\langle i,j \rangle} P_i^{\dag} P_j + h.c.
\label{bosehub}
\end{eqnarray}
where $t_B$ is the hopping energy of the boson. The phase diagram is
strongly dependent on the boson chemical potential, $\mu_B$,
that is given by
\begin{eqnarray} 
\mu_B= -2 \mu + E_{{\bf K}} \, .
\end{eqnarray} 
In the absence of disorder the superfluid phases are separated
from the Mott insulator phases by lines of second order phase transition.
At $t_B=0$ the Mott insulator phases extend in the range 
$n-1<\mu_B/U<n$ where $n$ is a positive integer that gives the 
number of bosons per site. Notice that because of condition (\ref{condition})
we must always have $\mu_B \leq 0$.
Therefore, the only allowed state for the bosons is the
Mott insulator with $n=0$, that is, the vacuum. 
For a finite value of $t_B$ there is
a critical value of the hopping above which the system becomes a superfluid. 
However, even a small amount of disorder suppresses the Mott 
insulator-superfluid transition by the creation of an insulating state. 
Disorder is unavoidable in these 
systems since the charge comes from counter-ions out of the CuO$_2$ planes. 
Thus, the conclusion is that if the bosons are decoupled from the stripes 
the Bose system is in the insulating Bose glass state.

\section{Stripe fermions}
\label{stfe}

As we have argued the single particle excitations at the regions of high
density of the CCS is due to the magnetic confinement \cite{sasha}.
In the absence of the CCS the holes can move freely with hopping energy
$t$ in both in the $x$ and $y$ directions and the Brillouin zone 
is defined for $-\pi/a < k_x,k_y \leq \pi/a$. In the presence of
the CCS the translation symmetry of the lattice perpendicular to the
ordering vector is broken. Suppose that the stripes are separated by
a distance $N a$. Then the new Brillouin zone is given by 
$-\pi/a < k_x \leq \pi/a$
and  $-\pi/(N a) < k_y \leq \pi/(N a)$. Thus the original band has to be
folded back into this zone and gaps open in the single particle
spectrum generating $N$ new bands.

When $N \gg 1$
the tunneling between stripes is suppressed because of the large
distance. A WKB estimate of the transverse tunneling energy 
using the confining potential (\ref{confine}) gives:
\begin{eqnarray}
t_{\perp}(N) \approx \frac{\rho_S(N)^{3/4} N^{3/4}}{m^{1/4} a^{1/2}}
e^{- N^{3/2} a \sqrt{2 m \rho_S(N)}}
\label{tpern}
\end{eqnarray}
where $m$ is the hole mass. Observe that for $N \gg 1$ (\ref{rhosn}) shows
that $\rho_S \approx J/4$ and therefore the stripe fermions have large
gaps in their spectrum. Moreover, 
$t_{\perp}$ is exponentially suppressed at large $N$ and intra-stripe
charge fluctuations are suppressed.
As $N$ is decreased the confining potential gets weaker and eventually
vanishes at $N_m$ given in (\ref{nm}). Close to $N_m$ the system is
essentially 2D and the confining potential $V_C(y)$ can
be treated as a perturbation of the atomic one and small gaps of order
$\rho_S(N)$ will open at $\pi/(N a)$. At this point, $N \to N_m^-$, 
the bandwidth of the lowest band is simply $4 t (1-\cos(\pi/N_m)) < 4 t$. 
Exactly at $N_m$ the stiffness
vanishes and the system becomes 2D since the holes can move 
freely, that is $t_{\perp}(N>N_m) = t$ where $t$ is the hopping energy
in the absence of the CCS. A simple and 
convenient way to parameterize the 
hopping in the transverse direction in the whole parameter range is
\begin{eqnarray}
t_{\perp}(N) = t e^{-(N-N_m)/N_0} \,
\label{tperpn} 
\end{eqnarray}
where $N_0 \approx N_m/\ln[1/(1-\cos(\pi/N_m))]$. 
The hopping along the stripes, $t_{||}$, 
is not so sensitive to the distance between stripes. It has been shown 
\cite{sasha} that for $N \to \infty$ the hopping energy is reduced 
from $t$ (as in the uniform system)
due to the dressing by strings to a value of
the order of $J \approx t$. On the other hand when $N \approx 1$ one 
has $t_{||} = t$. Thus, $t_{||}$ is
a smooth function and its variation with $N$ can be ignored. The anisotropy
in the hopping energies will lead to anisotropies in the Fermi surface 
as well. Suppose the stripes are oriented along the $x$ axis. 
The stripe fermion dispersion relation can be written as
\begin{eqnarray}
\epsilon_{{\bf k}}(N) = -2 t_{||} \cos(k_x a) - 2 t_{\perp}(N) \cos(N k_y a)
\label{dispersion2d}
\end{eqnarray}
and it is strongly anisotropic as long as $t_{||} \gg t_{\perp}$ (this is 
always true for $N>N_m$). In the limit of $N \gg 1$ the transverse
component can be completely disregarded and the Fermi surface is obtained
by filling up all the states up to $k_x = \pm k_F$ where $k_F$ is the
Fermi momentum (see Fig.\ref{bz}(a)). Moreover, the Fermi surface is open 
along $k_y$ since the system has no dispersion in that direction. 
If the stripes are oriented along the diagonals on a square lattice then
in momentum space the dispersion looks like the one given in Fig.\ref{bz}(b).
Then it
is convenient to rotate the orientation of the lattice by $\pi/4$ 
and to change the lattice spacing from $a$ to $a/\sqrt{2}$. 
Introduction of $t_{\perp}$ leads to a change in the curvature of
the Fermi surface by an amount $\delta k_{\perp}$ close to the Fermi
points.

As it has been noted in the context of organic superconductors \cite{jerome}
the strong anisotropy of the hopping produces strong dependence of the
physical properties with temperature. At high temperatures, that is,
$T > t_{\perp}(N)$, the uncertainty on the transverse momentum is larger
than the size of the Brillouin zone, $\delta k_{\perp} > \pi/(N a)$, and
the curvature of the Fermi surface is blurred by thermal effects. In this
case the motion of the electrons perpendicular to the stripes becomes
incoherent and electrons become confined by thermal effects to a region
of size $N a$. Thus, the motion of the electrons is essentially
1D and the 2D aspect of the stripe problem can be 
disregarded. In the opposite limit, $T < t_{\perp}(N)$, the curvature
of the Fermi surface is larger than the thermal effects and 2D
physics start to play a role. So there is a 1D to 2D
crossover as a function of temperature that occurs at $T_{cr}(N) 
\approx t_{\perp}(N)$.

In most of the paper we focus in the region where $N < N_m$
so it makes sense to talk about stripes and the existence of a CCS.
In this region the transverse hopping given by (\ref{tperpn}) is
exponentially small and the problem can be treated as purely 1D.
Consider a single isolated {\it static} a stripe that we describe 
as a 1D electron gas with dispersion $\epsilon_k$. The 
non-interacting Hamiltonian is simply
\begin{eqnarray}
H_0 = \sum_{k,\sigma} (\epsilon_k - \mu) \psi^{\dag}_{k,\sigma} \psi_{k,\sigma} \, .
\end{eqnarray}
As usual in the case of the 1D system we will assume that the physics
is dominated by the excitations close to the Fermi points.
Expanding the dispersion close to $\pm k_F$ ($\epsilon_{\pm k_F}=\mu$)
we can write the above Hamiltonian as
\begin{eqnarray}
H_0 = i v_F \sum_{\sigma} 
\int ds \, \left(\psi^{\dag}_{R,\sigma}(s) \partial_s \psi_{R,\sigma}(s)
-\psi^{\dag}_{L,\sigma}(s) \partial_s \psi_{L,\sigma}(s)\right)
\end{eqnarray}
where $s$ is the distance along the stripe (if the stripe is oriented along
the $x$ axis then $s=x$, but if the stripe is diagonal then $s=\sqrt{x^2+y^2}$)
$v_F$ is the Fermi velocity and  
$\psi_R(s)$ and $\psi_L(s)$ are
of right and left  moving electrons, respectively. The original fermion
operator is written as
\begin{eqnarray}
\psi_{\sigma}(s) = \psi_{R,\sigma}(s) e^{i k_F s}
+ \psi_{L,\sigma}(s) e^{-i k_F s}
\label{slowf}
\end{eqnarray}
and these can be bosonized via the transformation \cite{luttinger}
\begin{eqnarray}
\psi_{R,L,\sigma}(s) = \frac{1}{\sqrt{2 \pi a}} e^{\pm i \sqrt{\pi}
\phi_{R,L,\sigma}(s)} \, .
\nonumber
\end{eqnarray}
 
The bosonic modes
$\phi$ can be described in terms of amplitude, $\phi_{\sigma}$,
and phase, $\theta_{\sigma}$, bosonic modes as
$
\phi_{R,L,\sigma}(s) = \phi_{\sigma}(s) \mp \theta_{\sigma}(s) .
$
The bosonic fields can be rewritten in terms
of charge and spin bosonic modes:
$
\phi_{c,s} = \frac{1}{\sqrt{2}} \left(\phi_{\uparrow} \pm
\phi_{\downarrow}\right)$
and
$
\theta_{c,s} = \frac{1}{\sqrt{2}} \left(\theta_{\uparrow} \pm
\theta_{\downarrow}\right)
$.
If the average density on the stripe is $\rho_0$ then we can
write,
$\sum_{\sigma=\uparrow,\downarrow} \psi^{\dag}_{\sigma}(s) \psi_{\sigma}(s)=
a (\rho_0 + \delta \rho)$
and
$\delta \rho = \partial_s \phi_{\rho}/\sqrt{\pi}$
in bosonized form. The Hamiltonian of the problem for {\it forward}
scattering reads \cite{luttinger}:
\begin{eqnarray}
H_F = \sum_{\mu=\rho,s} \frac{v_{\mu}}{2} \int ds \left\{K_{\mu} \left(
\partial_s \theta_{\mu}\right)^2 + K^{-1}_{\mu} \left(
\partial_s \phi_{\mu}\right)^2 \right\}
\label{HF}
\end{eqnarray}
where $K_{\mu}$ are the Luttinger parameters (dependent on the 
electron-electron interactions) and
\begin{eqnarray}
v_{\mu} = \frac{v_F}{K_{\mu}}
\label{vmu}
\end{eqnarray}
are the charge and spin velocity. For the non-interacting system 
we have $K_{\rho}=K_{s}=1$. 
By the same token the action that describes the bosonized fermions 
is given by,
\begin{eqnarray}
S &=& \sum_{\mu} \frac{K_{\mu}}{2 v_{\mu}} \int ds \int d\tau
\left[\left(\partial_{\tau} \theta_{\mu}\right)^2 + v_{\mu}^2  
\left(\partial_{s} \theta_{\mu}\right)^2\right]
\nonumber
\\
&=& \sum_{\mu} \frac{1}{2 K_{\mu} v_{\mu}} \int ds \int d\tau
\left[\left(\partial_{\tau} \phi_{\mu}\right)^2 + v_{\mu}^2  
\left(\partial_{s} \phi_{\mu}\right)^2\right] \, .
\label{action1D}
\end{eqnarray}

Besides forward scattering we may also include a backscattering term:
\begin{eqnarray}
H_1 &\propto& \int ds \, \, \psi^{\dag}_{R,\uparrow}(s) \psi^{\dag}_{L,\downarrow}(s) 
\psi_{R,\downarrow}(s) \psi_{L,\uparrow}(s) + h.c.
\nonumber
\\
&=& g_1 \int ds \, \, \cos(\sqrt{8 \pi} \phi_s(s))
\label{h1}
\end{eqnarray} 
where $g_1$ is the coupling constant. In principle a Luttinger liquid
can also include Umklapp terms. The Umklapp terms are only important at commensurate
fillings and especially at half-filling when they are responsible for the Mott-Hubbard
gap. However, half-filling for the stripes implies that there is one electron
per site,
that is, the stripe is depopulated. This can only happen if the boson level crosses
the chemical potential and the stripe state becomes unstable. Throughout the paper
we assume that the stripe state is stable and therefore away from 
half-filling. Experiments by Tranquada and collaborators in
LNSCO indicate that the stripe filling is $1/4$ \cite{tranquada}. 
Thus, we disregard the Umklapp processes. 

The importance of the backscattering term can be readily understood by
a simple perturbative renormalization group (RG) calculation. 
It is a simple exercise 
to show that if we shrink the bandwidth of the stripe electrons
from $E_c$ to $E_c - dE_c$ the coupling constant $g_1$ renormalizes 
according to \cite{luttinger}
\begin{eqnarray}
\frac{\partial g_1}{\partial \ell} = 2 (1 - K_s) g_1(\ell)
\end{eqnarray}
where $\ell = \ln(1/E_c)$. Thus, $g_1$ is irrelevant for $K_s<1$ and relevant
for $K_s>1$. For repulsive interactions it turns out that $K_s<1$ and therefore
the interaction is irrelevant and the system is described by a 
gapless Luttinger liquid. 
If interactions are attractive $K_s>1$ and the backscattering is relevant
leading to the opening of a {\it spin gap} in the system signaling tendency to
superconducting fluctuations. We have started from a repulsive model and 
therefore
we assume all the interactions to be repulsive throughout the paper. 
Although the backscattering term is irrelevant it is well-known that the
slowest decaying correlation function for a repulsive interacting Luttinger 
liquid is the CDW one (it is equivalent to the case of $K_{\rho}<1$). 

If the CDW is
commensurate with the lattice (as it seems to be in the case of LSCO) then
the lattice pinning is strong and the system becomes insulating
\cite{gruner}. If the
CDW is incommensurate then a Luttinger liquid state is possible but any amount
of disorder will lead to CDW pinning driving the system again towards an
insulating regime. This is certainly undesirable from the phenomenological
point of view since these materials
are still metallic at finite temperatures. Many approaches that start
from a purely 1D description of the stripes are confronted
with this serious problem \cite{fradkin}. 
These approaches usually invoke 
the introduction of long-range interactions among the stripes. It is hard
to reconcile the long range interactions between different stripes with
the short range interacting within the stripes. For that to happen the
dielectric function of the material should be extremelly anisotropic (
experimentally it is known that this is not the case \cite{sridhar}).
We believe that once the long-range forces have done their work in
creating the stripe array the final metallic state screens any long-range
forces left, leading the system to purely local interactions. If the inter-stripe
interactions remain long-range the ground state should be a Wigner crystal \cite{branko} 
instead of a CDW and in both cases the stripes should be insulating. We stick with the
assumption that after the CCS is formed the interactions between
the elementary excitations discussed here are local. 
The ground state is a CDW state and therefore insulating. In fact this is
the picture that arises from mean-field approaches to the stripe problem
since in these mean-field studies the stripes are non-interacting
\cite{hartfock,ortiz}.

In the case of the CCS state where there are an infinite number of stripes
we add a new index to the fermion operator, $\psi_{n,\sigma}(s)$,
where $n$ labels the particular stripe. 
The Fourier transform of the electron
operator can be defined as
\begin{eqnarray}
\psi_{n,\sigma}(x) = \sqrt{N a} \int_{-\pi/a}^{\pi/a} \frac{dk_x}{2 \pi}
\int_{-\pi/(N a)}^{\pi/(N a)} \frac{dk_y}{2 \pi} e^{i (k_x x+k_y n N a)}
\psi_{\sigma,{\bf k}}
\label{ftrans}
\end{eqnarray}
where the anti-commutation relations are preserved ($\{\psi_{n,\sigma}(s),
\psi_{m,\sigma'}(x')\}=
\delta(x-x') \delta_{n,m} \delta_{\sigma,\sigma'}$ and 
$\{\psi_{\sigma,{\bf k}},\psi_{\sigma',{\bf k'}}\} = (2 \pi)^2 \delta({\bf k}
-{\bf k'})$).

\section{Coupling between stripe fermions and 2D bosons}
\label{coupl}

As we have discussed in the last two sections in the absence of 
coupling between
stripe fermions and bosons the ground state is insulating:
a Bose-Mott insulator in the antiferromagnetic 
regions and a CDW state along the
stripes. In order to determine the coupling between bosons and fermions we 
consider simple arguments based on conservation of momentum and energy.

As we have argued, the nature of the elementary excitations in the problem 
depends
strongly on their position in real space. On the one hand, bound pair bosons
are not eigenstates of the Luttinger liquid. 
On the other hand, Luttinger liquid bosons (collective sound
waves) are not elementary excitations of the doped Mott
insulator. Thus, the simplest process that couples these two types of 
excitations
is a decay process in which bosons are continuously transformed 
in stripe fermions and vice-versa.

\begin{figure}
\epsfysize6 cm
\hspace{0cm}
\epsfbox{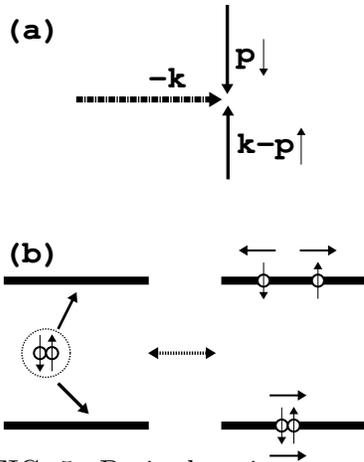}
\caption{Basic decaying processes for static stripes: (a) momentum exchange;
(b) real space process where a boson pair can decay into stripe fermions.}
\label{process}
\end{figure}

Let us consider the case of static stripes. The relevant coupling
that conserves linear momentum can be written as:
\begin{eqnarray}
V_{{\bf k}}({\bf q}) P_{-{\bf k}} 
\psi_{{\bf q},\uparrow} \psi_{{\bf k}-{\bf q},\downarrow} \, ,
\label{firstry}
\end{eqnarray}
where $V_{{\bf k}}({\bf q})$ is the boson-stripe coupling that, 
by conservation of angular momentum,
has d$_{x^2-y^2}$ symmetry (that is, $V_{{\bf k}}(0) \propto \cos(k_y a)-
\cos(k_x a)$).
This is a four-fermion coupling in terms of the original operators 
(see (\ref{pk})) and describes a fermionic scattering
process where in the final state two electrons always end up in
a bound state. This kind of coupling looks unusual
because there are only annihilation operators and it corresponds to the vertex
shown in Fig.\ref{process}(a). It only happens this way because we use
the electrons as assymptotic states in the scattering process.
In fact, (\ref{firstry}) is similar to
fermion-boson models that have appeared in the literature of superconductivity
over the years \cite{micnas,rumer}.
Notice that since the boson is composed of
holes the direction of the current is inverted (holes moving forward
are equivalent to electrons moving backward and vice-versa).
Thus to destroy
a boson is equivalent to create two electrons in the antiferromagnetic 
background. 
In the process (\ref{firstry}) a pair of holes is destroyed in the antiferromagnet
while a pair of electrons from the stripe hop into the antiferromagnetic media
(see Fig.\ref{process}(b)). 

Let us assume that the coupling between bosons and stripes
is weak so that the important degrees of freedom only exist close to
the chemical potential. Thus, when a boson hits a stripe there are two
main process for decaying. In the first one the two stripe fermions have
opposite momentum, that is, 
${\bf q} \approx k_F {\bf x}$ and ${\bf k}-{\bf q} \approx -k_F {\bf x}$
(as shown in Fig.\ref{process}(b)). Thus, linear momentum conservation 
implies ${\bf k}=0$. Moreover, angular momentum cannot be conserved 
in the scattering process because,
relative to the stripe, the two outgoing holes have zero angular momentum
while the pair have finite angular momentum. 
Therefore, we need a sink of momentum. The other process corresponds to boson 
decay into two fermion moving in the same direction,
that is, ${\bf q} \approx k_F {\bf x}$ and 
${\bf k}-{\bf q} \approx k_F {\bf x}$. This process
implies $k_x = 2 k_F$ and $k_y=0$. This is equivalent to a boson
moving in the antiferromagnetic ladder parallel to the stripe. Since the boson
wave-function is restricted to the antiferromagnetic region the amplitude
of this coupling is suppressed. The conclusion from this argument is that
for a static stripe the decay processes have vanishing phase space.

Consider also the difference between longitudinal and diagonal stripes as shown
in Fig.\ref{cou}. For a longitudinal stripe the coupling between bosons and
stripes is possible because the stripes are oriented along the lobes of the
boson d$_{x^2-y^2}$ 
wavefunction while in the diagonal case the stripes are oriented along
the nodes. Thus, the coupling between diagonal stripes is further reduced
as compared to the longitudinal case. Therefore, for diagonal stripes 
superconductivity is not possible in our model because of the nature of
the boson wave-function. In fact, recent neutron scattering experiments
observe diagonal stripes in the insulating phase of LSCO while vertical and
horizontal stripes are observed in the superconducting phase \cite{fujita}.
Moreover, notice that for a given orientation
of the boson wavefunction, horizontal (vertical) stripes couple to the
positive (negative) lobe of the wavefunction. 

As we pointed out above we need a sink (a particle or collective mode) 
for the momentum transferred from the bosons. Moreover, since the bosons
are spinless the particle that carries the momentum cannot be
a spin-wave or paramagnon. The simplest choice is a
vibrational mode: the boson collides with the stripe, breaks into two fermions
and produces a vibration. This vibration is due to the fluctuations
of the smectic phase of the stripe array. Again, as we
stressed previously, stripes are not isolated objects 
that are independent of each
other. Long-range forces, like the Coulomb, the Casimir or simply entropic
repulsion keeps them apart and generate a finite stiffness.

\begin{figure}
\epsfysize6 cm
\hspace{0cm}
\epsfbox{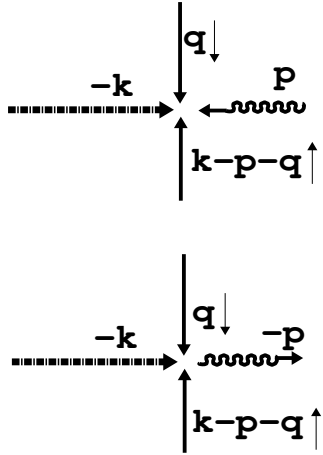}
\caption{Scattering processes including vibrations.}
\label{cpp}
\end{figure}

Consider 
the coupling of the bosons with the stripe fermions when the stripe fluctuates
a distance $u$ from its equilibrium position. We write the local
coupling as
\begin{eqnarray}
P({\bf r}) \psi_{\uparrow}({\bf r}) \psi_{\downarrow}({\bf r}) 
\delta(y - n N a - u(x,n N a)) 
\end{eqnarray}
where we ensure that the coupling occurs at the position of the stripes.
If we assume that CCS fluctuations are small compared to the 
periodicity in the system, that is, $|u| \ll N a$, 
the stripes fluctuate in a distance scale 
much smaller than the inter-stripe distance. We can expand the Dirac delta
function. The first term is just the static stripe problem we discussed above
and, as we argued, has no phase space. We therefore keep the
next term in the expansion leading to a coupling of the form
\begin{eqnarray}
- V_{{\bf k}}({\bf q}) \lambda_{{\bf p},\alpha} 
P_{-{\bf k}} \psi_{{\bf q},\uparrow} \psi_{{\bf k}-{\bf p}-{\bf q},\downarrow}
\left(b_{{\bf p,\alpha}} + b^{\dag}_{-{\bf p,\alpha}}\right)
\label{couplings}
\end{eqnarray}
where $\lambda_{{\bf p},\alpha} \propto p_y \to 0$ when $p \to 0$ 
is the stripe-vibration coupling constant. The two processes associated with
the interaction in (\ref{couplings}) are shown in Fig.\ref{cpp}.

\begin{figure}
\epsfysize6 cm
\hspace{0cm}
\epsfbox{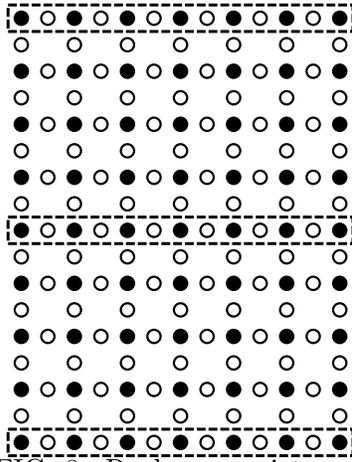}
\caption{Real space picture of the normal phase. Open circles: O atoms;
Filled circles: Cu atoms; dashed line: equilibrium position of the stripes.}
\label{abovetc}
\end{figure}

\begin{figure}
\epsfysize6 cm
\hspace{0cm}
\epsfbox{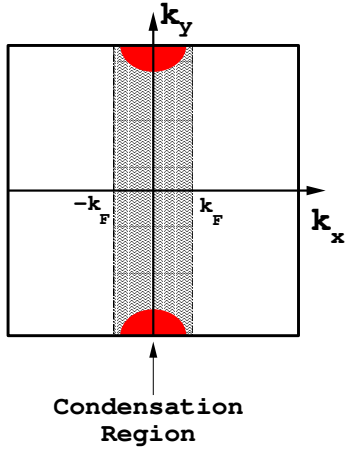}
\caption{Momentum space picture of the condensation region.}
\label{cond}
\end{figure}

Before we study (\ref{couplings}) in detail it is worth
investigating the problem via second order time dependent perturbation theory. 
The transition probability between non-perturbed states in the system is
given by:
\begin{eqnarray}
W(t) \approx \sum_{{\bf k},{\bf p},{\bf q},\alpha}
4 |V_{{\bf k}}({\bf q})|^2 |\lambda_{p,\alpha}|^2 
(1-\overline{n}_{{\bf q}}) (1-\overline{n}_{{\bf k}+{\bf p}-{\bf q}})
\frac{\sin^2[(-E_{{\bf k}} \pm \Omega_{{\bf p},\alpha}+\epsilon_{q_x}+
\epsilon_{k_x+p_x-q_x})t/2]}{
(-E_{{\bf k}}\pm \Omega_{{\bf p},\alpha}
+\epsilon_{q_x}+\epsilon_{k_x+p_x-q_x})^2}
\end{eqnarray}
where $\overline{n}_{{\bf q}}$ is the Fermi-Dirac distribution function. 
Observe that the probability amplitude only grows with time
for states such that
\begin{eqnarray}
-E_{{\bf k}}\pm \Omega_{{\bf p},\alpha}=-\epsilon_{q_x}-\epsilon_{k_x+p_x-q_x}
\label{energcons}
\end{eqnarray}
where the plus (minus) sign implies absorption (emission) of a vibration.
Remembering that $\epsilon_q \approx \mu$ we obtain the condition that 
$-E_{0,k_y}\pm \Omega_{0,k_y,\alpha}= -2 \mu$. 
Moreover, because $\lambda_{k_y}$ vanishes at $k_y \to 0$ the largest 
coupling between bosons and stripes happens
at {\it finite} momentum. Indeed, the coupling increases as one goes towards
the Brillouin zone edge. Thus, the largest coupling in the problem occurs
at ${\bf K} = (0,\pi/a)$, that is, {\it perpendicular} to the stripes.
It may be somewhat surprising that the most important part of the physics 
occurs in a direction perpendicular to the orientation of the stripes but
this effect has been already observed in numerical simulations of spin-fermion
models with stripe formation \cite{moreo}. 
This result
indicates that in the case of the boson condensation we should
expect it to occur close at ${\bf K}$ (as photo-emission experiments 
indicate \cite{arpes}) but not only that: we expect a CCS distortion at 
the same
wave-vector so that the momentum carried by the superfluid bosons is 
transfered to the lattice. Thus, there can be a double condensation: 
the superfluid density is modulated with finite wave-vector ${\bf K}$
and the CCS deforms in order to follow the variation of the superfluid
density. 
Therefore a condensation of the CCS at finite wave-vector will induce phonon anomalies
at the same wave-vector. Thus, we expect phonon softening close to $(0,\pi/a)$ together
with the condensation of bosons in the same point in momentum space. 
In the normal state the situation is
depicted in Fig.\ref{abovetc}: there is no superfluidity and no lattice distortions. 
In the ordered phase we expect the situation shown 
in Fig.\ref{belowtc} where the superfluid
density is modulated with wavelength $a/\pi$ in the direction perpendicular to the stripes
and the lattice is distorted in the same direction with the same wave-vector.
Notice that there is a doubling of the unit cell in the direction perpendicular
to the stripes and therefore the CCS is dimerized. This effect is similar to
the Peierls distortion in 1D systems or the Jahn-Teller effect 
where there is a lowering of the symmetry with a simultaneous gain in energy
\cite{jahn}.
In momentum space the situation is depicted in Fig.\ref{cond}: phonon anomalies and
superfluidity should appear at same point on the edge of the Brillouin zone. 

Observe that due to the conservation of energy,
$-E_{0,k_y}+2 \mu = \pm \Omega_{0,k_y,\alpha}$, and the stability condition (\ref{condition}) is
satisfied only if emission is allowed. One can generate a boson
by creating two electrons at the stripe at the cost of absorption of
a vibration. This can be clearly seen in
the diagram in Fig.\ref{dos} where in order to create a boson two
electrons occupying the boson state have to be excited to the Fermi surface
by the absorption of a vibration. Conversely, two holes from the stripe 
hop into the antiferromagnetic ladder and form a boson. 
Thus, our conclusion is that the most relevant coupling in this
problem is given by 
\begin{eqnarray}
H_c = - \sum_{{\bf k},{\bf p},{\bf q},\alpha} 
V_{{\bf k}}({\bf q}) \lambda_{{\bf p},\alpha} 
P_{{\bf k}} \psi_{{\bf q},\uparrow} \psi_{-{\bf k}+{\bf p}-{\bf q},\downarrow} 
b^{\dag}_{{\bf p},\alpha} + h.c. \, .
\end{eqnarray}

\section{Mean-Field Theory}
\label{mfield}

In order to understand the phase diagram of this problem it is illuminating
to consider the mean-field solution. Let us rewrite the full
Hamiltonian by collecting all the terms given in the previous sections 
(we neglect the interactions between 
electrons in the stripe for the moment being). The Hamiltonian is written as
\begin{eqnarray}
H &=& \sum_{{\bf k}} 
\left[(-E_{{\bf k}}+2 \mu) P^{\dag}_{{\bf k}} P_{{\bf k}}
+\Omega_{{\bf k}} b^{\dag}_{{\bf k}} b_{{\bf k}}\right] + 
\sum_{{\bf k},\sigma} (\epsilon_{{\bf k}}-\mu) 
\psi^{\dag}_{{\bf k},\sigma} \psi_{{\bf k},\sigma}
\nonumber
\\
&+& \frac{U}{2} \sum_i N_i^2 - 
\sum_{{\bf k},{\bf p},{\bf q}} V_{{\bf k}}({\bf q}) \lambda_{{\bf p}} P_{{\bf k}} 
\psi_{{\bf q},\uparrow} \psi_{-{\bf k}+{\bf p}-{\bf q},\downarrow} b^{\dag}_{{\bf p}} + 
h.c. \, 
\label{finalh}
\end{eqnarray}
where we have simplified the problem by considering a single polarization of
the vibration field perpendicular to the stripes.

Notice that the Hamiltonian is invariant under the gauge transformation
\begin{eqnarray}
P &\to& P \, \, e^{-i \varphi_B}
\nonumber
\\
b &\to& b \, \, e^{+i \varphi_V}
\nonumber
\\
\psi &\to& \psi \, \, e^{+ i (\varphi_B+\varphi_V)/2}
\label{gaugetrans}
\end{eqnarray}
implying the phase of the bosons and vibrations are coupled 
through the fermionic degrees of freedom. 
This symmetry is the one that is broken in the ordered state.

In the mean-field theory we split the coupling term into three different
pieces leading to a mean-field Hamiltonian of the form: $H_{MF} = 
H_P + H_B + H_S$, where:
\begin{eqnarray}
H_P &=& \sum_{{\bf k}} \left[
(-E_{{\bf k}}+2 \mu) P^{\dag}_{{\bf k}} P_{{\bf k}} - g_{P,{\bf k}}  P_{{\bf k}}
- g^*_{P,{\bf k}}  P^{\dag}_{{\bf k}}\right]
+  \frac{U}{2} \sum_i N_i^2 
\nonumber
\\
H_B &=&  \sum_{{\bf k}} \left[\Omega_{{\bf k}} b^{\dag}_{{\bf k}} b_{{\bf k}}
- g_{B,{\bf k}} b_{{\bf k}} - g^*_{B,{\bf k}} b^{\dag}_{{\bf k}}\right]
\nonumber
\\
H_S &=& \sum_{{\bf k},\sigma} (\epsilon_{{\bf k}}-\mu) 
\psi^{\dag}_{{\bf k},\sigma} \psi_{{\bf k},\sigma} 
- \sum_{{\bf k},{\bf p},{\bf q}} g_{S}({\bf k},{\bf p},{\bf q})
\psi_{{\bf q},\uparrow} \psi_{-{\bf k}+{\bf p}-{\bf q},\downarrow} + h.c.
\label{mfpi}
\end{eqnarray}
where
\begin{eqnarray}
g_{P,{\bf k}}  &=& \sum_{{\bf p},{\bf q}} V_{{\bf k}}({\bf q}) \lambda_{{\bf p}}
\langle \psi_{{\bf q},\uparrow} \psi_{-{\bf k}+{\bf p}-{\bf q},\downarrow} 
\rangle \, \langle b_{{\bf p}} \rangle
\nonumber
\\
g_{B,{\bf k}}  &=& \lambda_{{\bf k}} \sum_{{\bf p},{\bf q}}  V_{{\bf p}}({\bf q}) 
\langle P_{{\bf p}} \rangle \, \langle \psi_{{\bf q},\uparrow} 
\psi_{{\bf p}+{\bf k}-{\bf q},\downarrow} \rangle
\nonumber
\\
g_{S}({\bf k},{\bf p},{\bf q}) &=& V_{{\bf k}}({\bf q}) \lambda_{{\bf p}} 
\langle P_{{\bf k}} \rangle \, \langle b_{{\bf p}} \rangle
\end{eqnarray}
are the mean-field coupling constants. From (\ref{mfpi}) it becomes obvious
the symmetry breaking processes in the ordered phase: 
spontaneous gauge symmetry breaking leading to 
superfluidity of the bosons, superconductivity of the fermions
and spontaneous symmetry breaking of translational symmetry of the
CCS.   

Thus, for a transition to a non-trivial conducting state 
the coupling $\lambda V$ is fundamental.
The simplest conducting state is the superfluid state where the charged
bosons condense into a macroscopic condensate. 
In accordance with the discussion
in the previous section, conservation rules tell us that the largest coupling
occurs at the largest momentum perpendicular to the stripes. 
Suppose the bosons condense at ${\bf k}={\bf K}$, that is, 
$\langle P_{{\bf K}}\rangle= \langle P^{\dag}_{{\bf K}}\rangle
= \sqrt{N_B}$. In order to conserve momentum in the interaction in 
(\ref{finalh}) and preserve time reversal symmetry one should also require the 
vibrations to condense, that is, $\langle  b^{\dag}_{{\bf K}} \rangle = 
\langle  b_{{\bf K}} \rangle = \sqrt{N_V}$. Notice that the 
Cooper pairs have zero center of mass momentum. 
This is equivalent to a static distortion of the lattice with modulation 
${\bf K}$. Thus, at the mean-field level, a transition to a superfluid state 
is accompanied by a static lattice distortion. We notice here that the choice
of the largest coupling as the one that drives the system into the ordered
state is the traditional one in mean-field theories. In fact, the doubling
the unit cell proposed here also occurs in some large $N$ treatments of the 
$t-J$ models without phonons \cite{subirnick,vojtasubir}. Thus, this
tendency of lowering the energy of the system by breaking the lattice symmetry
seems to be quite generic.

The solution of the bosonic problem is straightforward because it only
requires a shift of the boson operators by a constant:
\begin{eqnarray}
P_{{\bf K}} &\to& P_{{\bf K}} - \frac{g_{P,{\bf K}}}{-E_{{\bf K}}+2 \mu}
\nonumber
\\
b_{{\bf K}} &\to& b_{{\bf K}} - \frac{g_{B,{\bf K}}}{\Omega_{{\bf K}}}
\end{eqnarray}
implying that the actual CCS distortion is given by
\begin{eqnarray}
\delta u({\bf K}) = \frac{\sqrt{2} \Re[g_{B,{\bf K}}]}{M_V^{1/2} 
\Omega^{3/2}_{{\bf K}}} \, .
\label{du}
\end{eqnarray}
where $M_V$ is the mass associated with the vibrations. 

The electronic problem can be also readily solved since it is a problem of
electrons with a pairing potential. We can write $H_S$ as
\begin{eqnarray}
H_S=\sum_{{\bf k}} \left[(\epsilon_{{\bf k}}-\mu) \sum_{\sigma} 
\psi^{\dag}_{{\bf k},\sigma} \psi_{{\bf k},\sigma} 
- \Delta_{{\bf k}} \psi_{{\bf k},\uparrow} \psi_{-{\bf k},\downarrow} + h.c.\right]
\label{hs}
\end{eqnarray}
where the superconducting gap is given by
\begin{eqnarray}
\Delta_{{\bf k}} = \lambda_{{\bf K}} a^2 \sqrt{\sigma_B^0 \sigma_V^0} \, \, V_{{\bf k}}(0)
\label{gap}
\end{eqnarray}
and therefore has d$_{x^2-y^2}$ symmetry as the pairs that generate it in the 
first place. Here we have defined 
\begin{eqnarray}
\sigma_B^0 &=& \frac{N_B}{S}
\nonumber
\\
\sigma_V^0 &=& \frac{N_V}{S}
\end{eqnarray}
as the superfluid and distortion densities, respectively.
The Hamiltonian (\ref{hs}) can be diagonalized by a Bogoliubov transformation:
\begin{eqnarray}
\psi_{{\bf k},\uparrow} &=& \cos(\theta_{{\bf k}}) d_{{\bf k},\uparrow} 
+ \sin(\theta_{{\bf k}}) d^{\dag}_{{\bf k},\downarrow}
\nonumber
\\
\psi^{\dag}_{-{\bf k},\uparrow} &=& \cos(\theta_{{\bf k}}) d^{\dag}_{{\bf k},\downarrow} 
- \sin(\theta_{{\bf k}}) d_{{\bf k},\uparrow}
\end{eqnarray}
where
\begin{eqnarray}
\tan(2 \theta_{{\bf k}}) = \frac{\Delta_{{\bf k}}}{\epsilon_{{\bf k}}-\mu}
\end{eqnarray}
and
\begin{eqnarray}
H_S = \sum_{{\bf k}} e_{{\bf k}} d^{\dag}_{\sigma,{\bf k}}d_{\sigma,{\bf k}}
\end{eqnarray}
where
\begin{eqnarray}
e_{{\bf k}} = \sqrt{(\epsilon_{{\bf k}}-\mu)^2+|\Delta_{{\bf k}}|^2}
\end{eqnarray}
is the quasiparticle dispersion.

The free energy per unit of area, $f(T)$, in the system can be obtained in
a straightforward way:
\begin{eqnarray}
f(T) = (-E_{{\bf K}} + 2 \mu) \sigma_B^0 + \frac{U}{2} (\sigma_B^0)^2 
+ \Omega_{{\bf K}} \sigma_V^0 
+ \frac{1}{S} 
\sum_{{\bf k}} \left[(\epsilon_{{\bf k}}-\mu)
- e_{{\bf k}} (1-2 n_{{\bf k}})\right]
\label{ksmf}
\end{eqnarray}
where
\begin{eqnarray}
n_{{\bf k}} = \frac{1}{e^{\beta e_{{\bf k}}}+1}
\label{nek}
\end{eqnarray}
is the quasiparticle occupation. 
We minimize (\ref{ksmf}) 
keeping the number of particles fixed. We obtain two equations:
\begin{eqnarray}
-E_{{\bf K}}+ 2 \mu + U \sigma_B^0 &=& \frac{\lambda^2_{{\bf K}} a^2 \sigma_V^0}{2 S} 
\sum_{{\bf k}} 
\frac{V^2_{{\bf k}}}{e_{{\bf k}}}
tanh(\beta e_{{\bf k}}/2)
\nonumber
\\
\Omega_{{\bf K}} &=&  
\frac{\lambda^2_{{\bf K}} a^2 \sigma_B^0}{2 S} \sum_{{\bf k}} 
\frac{V^2_{{\bf k}}}{e_{{\bf k}}}
tanh(\beta e_{{\bf k}}/2)
\end{eqnarray}
and therefore
\begin{eqnarray}
\Omega_{{\bf K}} \sigma_V^0 &=& \sigma_B^0 (-E_{{\bf K}}+2 \mu + U \sigma_B^0)
\nonumber
\\
&\approx&  U (\sigma_B^0)^2
\label{ratiovb}
\end{eqnarray}
that is, the superfluid and vibration density are tied to each other.
The last line in (\ref{ratiovb}) comes from the fact that we expect
$U \gg -E_{{\bf K}}+2 \mu$.
Using the above equation we reduce the mean-field problem to a single equation
\begin{eqnarray}
\Omega_{{\bf K}} = \frac{\lambda^2 a^2 \sigma_B^0}{2 S} \sum_{{\bf k}} \frac{V^2_{{\bf k}}}{e_{{\bf k}}}
tanh(\beta e_{{\bf k}}/2) \, .
\label{gapeq}
\end{eqnarray}

Notice that although the coupling constant has d-wave symmetry the Fermi
surface of an infinite array of 1D stripes has no curvature (as shown in
Fig.\ref{bz}(a)) because the fermions do not propagate in the direction
perpendicular to the stripes. Thus, although the order parameter in (\ref{gap})
has d$_{x^2-y^2}$ symmetry, the Fermi surface is fully gapped.
The bosons have a tendency to produce a d-wave superconductor but
the condensate only takes advantage of one of the lobes of the boson
wave-function. Thus, the final result is an s-wave state. As we will 
discuss in section \ref{conclu} 
a true d-wave superconductor can only develop if
the system is made out of domains with horizontal and vertical stripes
or a one particle tunneling is included (changing the fermion 
dispersion relation as in (\ref{dispersion2d})) and the
Fermi surface becomes rounded leaving space for a true d-wave state 
(although rather anisotropic).
This seems to indeed be the case of these systems where microscopic twin
boundaries are unavoidable. Because the domains are large the
superconducting properties of the domains can be studied as if the domains
are macroscopic superconductors. Taking the geometry of the Fermi surface
into account 
we can simplify the above equation by using the fact that 
(\ref{nek}) only depends on $k_x$ and therefore 
the sum over $k_y$ just gives $N_s/N$ leading to a simpler equation
\begin{eqnarray}
\Omega_{{\bf K}} = \frac{\lambda^2 V_0^2 a \sigma_B^0}{N} \int_{0}^{\pi/a} \frac{dk}{2 \pi} \frac{\tanh(\beta e_k/2)}{e_k}
\label{gapeq2}
\end{eqnarray}
where $V_0$ is the strength of the coupling $V_{\bf k}({\bf q})$ at the
stripe and $e_k = \sqrt{(\epsilon_{k}-\mu)^2+V_0^2 \lambda^2 (\sigma_B^0)^2 (-E_{{\bf K}}+2 \mu + U \sigma_B^0)/\Omega_{{\bf K}}}$.
This equation can be solved numerically assuming $\epsilon_{k_x} = - t
\cos(k_x a)$. 

\begin{figure}
\epsfysize10 cm
\epsfxsize10 cm
\hspace{0cm}
\epsfbox{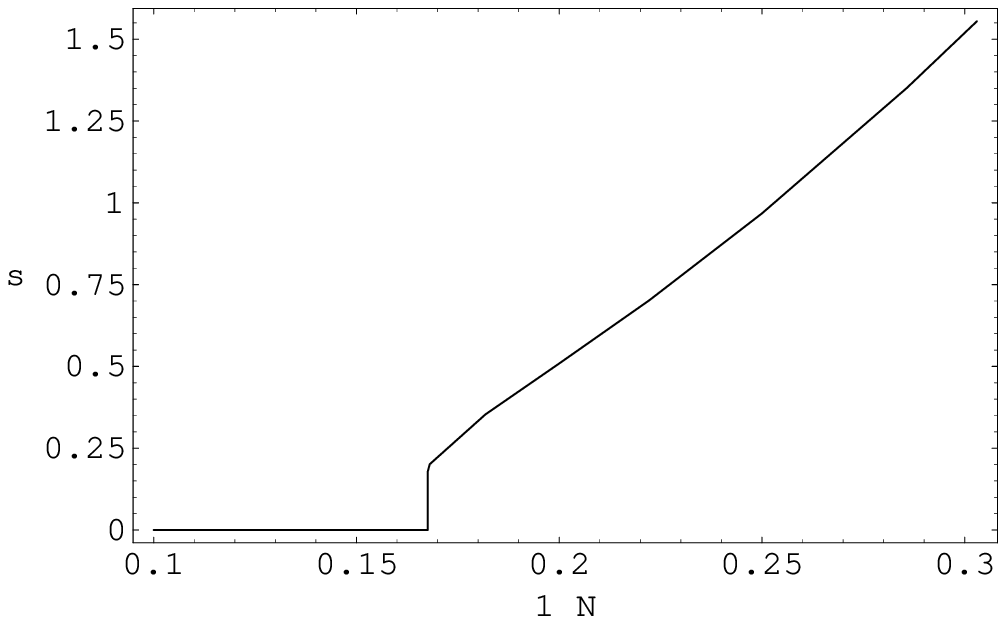}
\caption{$\sigma^0_B$ as a function of $1/N$.}
\label{sfd}
\end{figure}

\begin{figure}
\epsfysize10 cm
\epsfxsize10 cm
\hspace{0cm}
\epsfbox{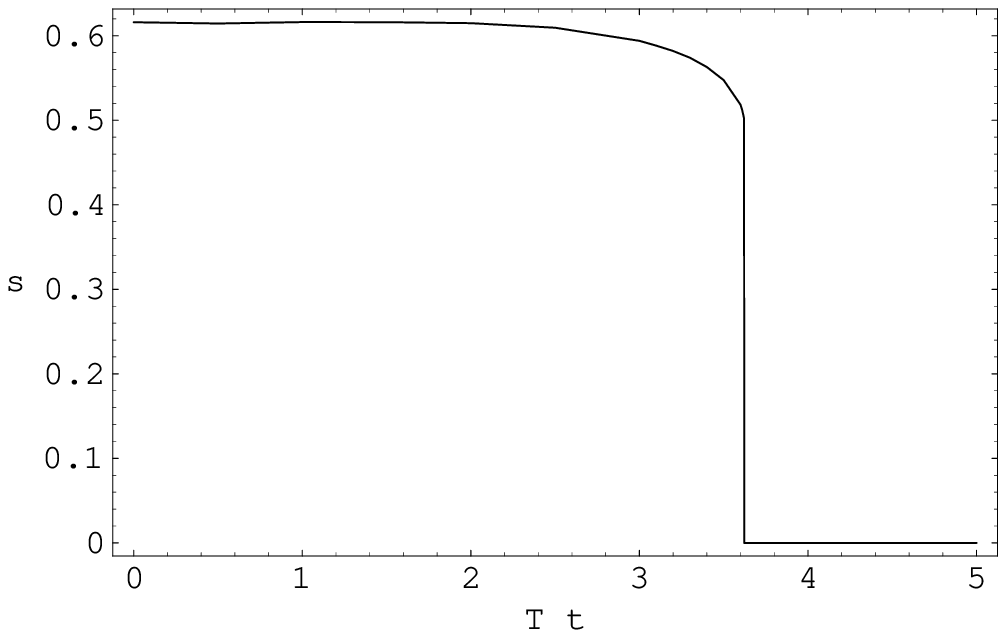}
\caption{$\sigma^0_B$ as a function of $T/t$.}
\label{sfdt}
\end{figure}

We can now study the phase diagram as a function of the coupling $\lambda V$
and the distance between stripes $N$. We expect 
the characteristic energy scales in the problem to be dependent
on the distance between stripes. This is certainly true for the case of 
antiferromagnetic order where the distance between the stripes determines
the strength of the quantum fluctuations \cite{hone}. Thus we expect the
phonon and boson spectra to change with $N$, that is, there must be
a continous change in the spectral weight with the interstripe distance.
In this paper, for simplicity, 
we will assume that this change is small and the
density of states does not change with $N$ (see, however, \cite{moreo}). 
In Fig.\ref{sfd} we plot 
$\sigma^0_B$ as a function of $1/N$ for $E_{{\bf K}} = 1.9 t$,
$\mu = -0.6 t$, $\Omega_{{\bf K}} = 1.2 \times 10^{-3} t$, $U =  4 t$,
$\lambda V = 0.5 t$ at $T=0$. Notice that there is a discontinuous jump
in $\sigma^0_B$ at $N_{sp} \approx 6$ and it grows in a quasi-linear
way with $1/N$. In Fig.\ref{sfdt} we plot $\sigma^0_B$ as a
function of $T/t$ for the same parameters given above. 
Notice that $\sigma^0_B$ is weakly dependent on the temperature
up to $T=T_{sp} \approx 3.5 t$ 
where it has a discontinuous jump to zero. As one can
see from the numerical solution $T_{sp}$ is of the order of
the bandwidth of the stripe fermions and therefore very large. As we show
below this is not the actual transition temperature because we have
not included phase fluctuations into the problem. Thus $T_{sp}$
(like $T^*$ defined in Section \ref{intro}) is a crossover temperature
scale. We note that $T_{sp} \gg T^*$, that is the mean-field
critical temperature is larger than the temperature scale above which
makes sense to talk about boson pairs and confinement. Therefore,
$T_{sp}$ is actually not observable since the theory brakes down at $T^*$.

\begin{figure}
\epsfysize10 cm
\epsfxsize10 cm
\hspace{0cm}
\epsfbox{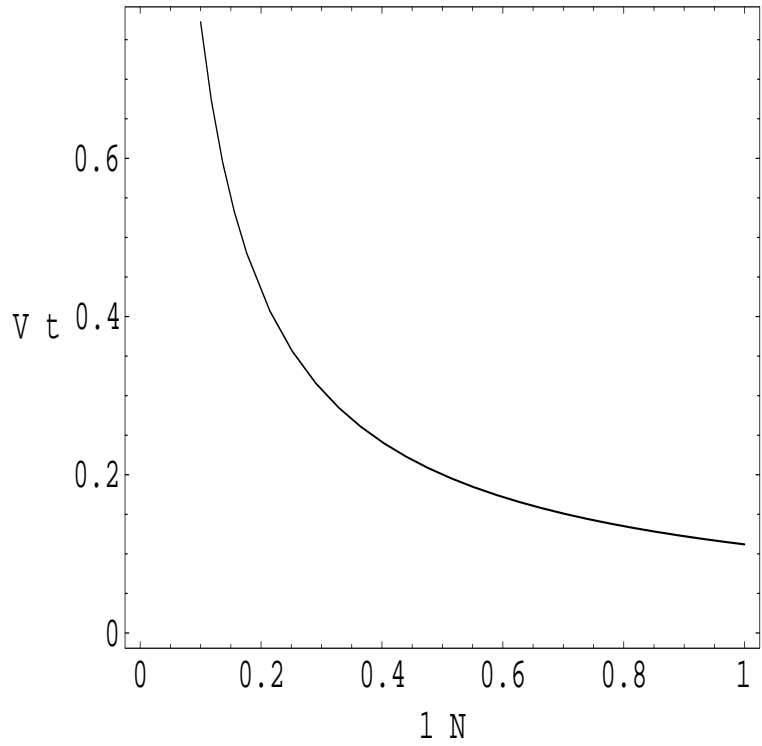}
\caption{Critical value of the coupling constant $V$ as a function of $1/N$.}
\label{vc}
\end{figure}

In order to understand the nature of the phase diagram we simplify the
problem by the 
introduction of a cut-off $\Lambda$ of the order of the inverse of the lattice spacing and 
linearize the spectrum at the chemical potential. The
integral simplifies and at $T=0$ reduces to:
\begin{eqnarray}
\Omega_{{\bf K}} = \frac{\lambda^2 V_0^2 a \sigma_B^0}{2 \pi v_F N} 
\ln\left(\frac{\Lambda v_F + \sqrt{(\Lambda v_F)^2 
+ \Delta_0^2}}{\Delta_0}\right)
\label{gapsimple}
\end{eqnarray}
where $\Delta_0 = V_0 \lambda \sigma_B^0 \sqrt{(-E_{{\bf K}} +2 \mu + U \sigma_B^0)/
\Omega_{{\bf K}} }$ gives the amplitude of the gap.
Notice that the transition at the mean-field level 
is of first order because the
r.h.s. of (\ref{gapsimple}) vanishes as $\sigma^0_B \to 0$ in
contrast with the usual BCS equation that contains only the logarithm
and is finite in this limit. 
Some analytical progress can be made in the limit of large $U$. 
The critical value of $N$ and the value 
of the superfluid fraction are given by:
\begin{eqnarray}
N_{sp} &\approx& \frac{\ln(1+\sqrt{2})}{2} \frac{(\lambda V_0)^{4/3}}{(\pi v_F/a)^{1/3} \Omega_{{\bf K}}^{2/3} U^{1/3}}
\nonumber
\\
\sigma_{0,c} &\approx& \frac{\Omega_{{\bf K}}^{1/3}(\pi v_F/a)^{2/3} }{
(\lambda V_0)^{2/3} U^{1/3}}
\end{eqnarray}
that determines the zero temperature mean-field phase diagram. As we
vary $N$ there is a critical value of $V_0$ (that we call $V_c$) below
which order is not possible. In Fig.\ref{vc} we plot $V_c$ as a function
of $1/N$. We find that $V_c \propto N^{\alpha}$ with $\alpha>0$. 
This result 
demonstrates that as the stripes get further apart a larger coupling
is required to stabilize long-range order in the system.

As pointed out by EK, the fact that the
superfluid density is so low implies large phase fluctuations 
\cite{emerykivelson}.
If one includes
the phase of the order parameters (as in (\ref{gaugetrans})) 
long-range order requires
that
\begin{eqnarray}
\langle P_{{\bf K}} \rangle &=& \langle |P_{{\bf K}}| \rangle \, \langle e^{-i \varphi_B} \rangle
\neq 0
\nonumber
\\
\langle b_{{\bf K}} \rangle &=& \langle |b_{{\bf K}}| \rangle \, \langle e^{+i \varphi_V} \rangle
\neq 0
\end{eqnarray}
where $\langle |P_{{\bf K}}| \rangle \propto \sqrt{\sigma_B^0}$ and
$\langle |b_{{\bf K}}| \rangle \propto \sqrt{\sigma_V^0}$ are the amplitudes of the order
parameters. Thus, ordering also requires 
$\langle e^{\pm i \varphi_{\alpha}} \rangle \neq 0$
with $\alpha=B,V$. In our mean-field theory we have fixed arbitrarily 
$\varphi_{\alpha}=0$. This is not 
correct since fluctuations of $\varphi_{\alpha}$, especially in
the 2D system, are fundamental for the development of long-range order.
In the next sections we discuss the role of these fluctuations in
determining the actual phase diagram. 
In other words, at high temperatures bosons 
and superconducting stripe fermions are formed but they are incoherent 
because of the existence
of topological defects: vortex-antivortex pairs in the superfluid and 
dislocation loops of the distorted
CCS.
The interesting result of our calculation is that although the amplitudes of the
two order parameters appear at the same in temperature $T^*$
(as given by (\ref{ratiovb}))
the actual ordering temperatures can be different because they depend on the
phase stiffness of the relevant degrees of freedom. 

\section{Phase fluctuations}
\label{phflu}

As we discussed at the end of the previous section the problem of 
phase fluctuations
is fundamental for the description of the ordered state. Therefore
the topological excitations will be fundamental for the determination
of the properties of the superconducting state \cite{jan}.
The simplest way to
discuss the importance of phase fluctuations is by studying the 
partition function
of the problem that can be written as:
\begin{eqnarray}
Z = \int D\overline{P} D P D\overline{\psi} D\psi D\overline{b} D b \, \, \, 
e^{-\int_0^{\beta} d\tau L[\overline{P},P,\overline{\psi},\psi,\overline{b},b]}
\end{eqnarray}
where
\begin{eqnarray}
L &=& \int d^2 r \left\{ \overline{P}({\bf r}) 
(\partial_{\tau}+E(\nabla)+2\mu) P({\bf r}) + 
\frac{U}{2} N^2({\bf r})
+ \overline{b}({\bf r}) (\partial_{\tau}+\omega(\nabla)) b({\bf r})\right\}
\nonumber
\\
&+& \sum_{n}
\left\{ \int dx \sum_{\sigma} \overline{\psi}_{n,\sigma}(x) 
(\partial_{\tau}+\epsilon(\nabla)-\mu) \psi_{n,\sigma}(x)
\right.
\nonumber
\\
&+& \left. \int dx  \int dx' \sum_{\sigma,\sigma'} 
\overline{\psi}_{n,\sigma}(x) \psi_{n,\sigma}(x) W_{\sigma,\sigma'}(x-x') 
\overline{\psi}_{n,\sigma'}(x') \psi_{n,\sigma'}(x') \right\}
\nonumber
\\
&-&\sum_{n}\int dx V(x,n N a) \left\{P(x,n N a) \psi_{n,\uparrow}(x) 
\psi_{n,\downarrow}(x) \overline{b}(x,n N a) + c.c.\right\}
\label{lag}
\end{eqnarray}
is the Lagrangian associated with Hamiltonian (\ref{finalh}) and $\beta$ is the inverse
temperature. 
Here we have redefined $\lambda V \to V$ and introduced the electron-electron
interactions $W_{\sigma,\sigma'}(x-x')$.

We redefine the fields in order to separate the problem into slow and fast variables:
\begin{eqnarray}
P({\bf r}) &=&\frac{1}{\sqrt{2}} \left( P_{+}({\bf r}) e^{-i {\bf K} \cdot {\bf r}} + P_{-}({\bf r}) e^{i {\bf K} \cdot {\bf r}}\right)
\nonumber
\\
b({\bf r}) &=& \frac{1}{\sqrt{2}} \left( b_{+}({\bf r}) e^{-i {\bf K} \cdot {\bf r}} + b_{-}({\bf r}) e^{i {\bf K} \cdot {\bf r}}\right)
\label{fastslow}
\end{eqnarray}
and perform the same expansion as in the case of fermions (see (\ref{slowf})).

Substituting (\ref{slowf}) and (\ref{fastslow}) into the Lagrangian (\ref{lag}) 
we find many terms that oscillate fast. We disregards all the oscillating
terms and study slow modes only. The Lagrangian can be
written as
\begin{eqnarray}
L = \sum_{\alpha=\pm} L_0[P_{\alpha},b_{\alpha}] + \sum_n L_S[\psi_n] + L_C
\end{eqnarray}
where 
\begin{eqnarray}
L_0[P_{\alpha},b_{\alpha}] &=& \frac{1}{2} \int d^2 r 
\left\{\overline{P}_{\alpha}({\bf r}) \left(\partial_{\tau}+E_0-\frac{\nabla^2}{2 M_B}\right) P_{\alpha}({\bf r}) 
+ \frac{U}{2} N^2 ({\bf r}) \right.
\nonumber
\\
&+& \left. \overline{b}_{\alpha}({\bf r}) \left(\partial_{\tau}+\omega_0-\frac{\nabla^2}{2 M_V}\right) b_{\alpha}({\bf r})\right\}
\end{eqnarray}
where we have assumed that ${\bf K}$ is a {\it local} minimum in
the energy of the bosons and vibrations. This assumption is supported
by studies of the $t-J$ model \cite{sushkov} and from the fact that the 
vibration mode is at the edge of the Brillouin zone.
Here, $M_B$ is the effective boson mass.
The form of the  non-interacting boson-vibration Lagrangian given
above is required by Galilean invariance of the problem in the long
wavelength limit. Although anisotropies in the dispersion of the modes
close to ${\bf K}$ should exist, 
we disregard them since they are not fundamental in our discussion.
Furthermore,
\begin{eqnarray}
L_S[\psi_n] &=& \int \frac{dx}{a} 
\left\{ \sum_{\sigma} \left[\overline{\psi}_{n,R,\sigma}(x)\left(\partial_{\tau}+i v_F \partial_x\right)
\psi_{n,R,\sigma}(x) + \overline{\psi}_{n,L,\sigma}(x)\left(\partial_{\tau}-i v_F \partial_x\right)\psi_{n,L,\sigma}(x)\right]
\right.
\nonumber
\\
&+& \left. \sum_{\sigma,\sigma'} \left[W_{k=0} (\overline{\psi}_{n,R,\sigma}(x) \psi_{n,R,\sigma}(x)
+\overline{\psi}_{n,L,\sigma}(x) \psi_{n,L,\sigma}(x)) \times
\right. \right.
\nonumber
\\
&\times& \left. \left. (\overline{\psi}_{R,\sigma'}(x,n) \psi_{R,\sigma'}(x,n)+
\overline{\psi}_{L,\sigma'}(x,n) \psi_{L,\sigma'}(x,n)) 
\right] \right. 
\nonumber
\\
&+& \left. \left[W_{k=2 k_F} (\overline{\psi}_{n,R,\sigma}(x) \psi_{n,L,\sigma}(x)
+\overline{\psi}_{n,L,\sigma}(x) \psi_{n,R,\sigma}(x)) \times \right. \right.
\nonumber
\\
&\times& \left. \left. 
(\overline{\psi}_{R,\sigma'}(x,n) \psi_{L,\sigma'}(x,n)+\overline{\psi}_{L,\sigma'}(x,n) \psi_{R,\sigma'}(x,n)) 
\right] \right\}
\end{eqnarray}
is the isolated stripe Lagrangian ($W_{{\bf k}}$ is the Fourier transform of the
electron-electron interaction) and
\begin{eqnarray}
L_C &=&  - \frac{1}{2} 
\sum_{n} \int dx V(x,n) \left[P_{+}(x,n) \overline{b}_+(x,n)
+P_-(x,n) \overline{b}_-(x,n)\right]
\left[\psi_{n,R,\uparrow}(x) \psi_{n,L,\downarrow}(x) 
\right. 
\nonumber
\\
&+& \left. \left. \psi_{n,L,\downarrow}(x) \psi_{n,R,\uparrow}(x)\right]
+ c.c. \right\}
\end{eqnarray}
is the boson-stripe-vibration coupling. 

Due to the symmetry we simplify the problem by changing to amplitude-phase
modes via:
\begin{eqnarray}
P_+ &=& P_- = \sqrt{\sigma_B} e^{-i \varphi_B}
\nonumber
\\
b_+ &=& b_- = \sqrt{\sigma_V} e^{+i \varphi_V}
\end{eqnarray}
and use the bosonization technique described previously so that
\begin{eqnarray}
L = \sum_n L_{1D}[\theta_{n,\rho},\phi_{n,s}] + \sum_{\alpha=B,V} L_{\alpha}
[\sigma_{\alpha},\varphi_{\alpha}] + L_C[\Phi,\varphi,\theta_{\rho},\phi_{s}]
\end{eqnarray}
where $L_{1D}[\theta_{n,\rho},\phi_{n,s}]$ is given in (\ref{action1D}).
Furthermore, 
\begin{eqnarray}
L_{\alpha} = \int d^2r \left[i \sigma_{\alpha} \partial_{\tau} \varphi_{\alpha}+ E_{\alpha} \sigma_{\alpha} + \frac{U_{\alpha}}{2} \sigma^2_{\alpha}
 + \frac{\sigma_{\alpha}^{-1}}{8 M_{\alpha}}
(\nabla \sigma_{\alpha})^2 + \frac{\sigma_{\alpha}}{2 M_{\alpha}}
(\nabla \varphi_{\alpha})^2\right]
\end{eqnarray}
with $E_B = E_{{\bf K}}$, $E_V = \Omega_{{\bf K}}$, 
$U_B = U$ and $U_V=0$ and,
\begin{eqnarray}
L_C = - \sum_n \int \frac{dx}{\pi a} V(x,n N a)
\sqrt{\sigma_B \sigma_V} e^{-i (\varphi_B(x,n N a) + \varphi_V(x,n N a) - \sqrt{2 \pi} 
\theta_{n,\rho}(x))} \cos(\sqrt{2 \pi} \phi_{n,s}(x)) + c.c. \, .
\end{eqnarray}

Observe that the gauge symmetry discussed in the previous section is explicit
in the Lagrangean. In order to gauge away the phase fields we 
define a new bosonic field $\theta_c$:
\begin{eqnarray}
\theta_{n,c}(x) = \theta_{n,\rho}(x) - \frac{\varphi_B(x,n N a)+\varphi_V(x,n N a)}
{\sqrt{2 \pi}}
\label{thetac}
\end{eqnarray}
so that $L_C$ simplifies to
\begin{eqnarray}
L_C = - \sum_n \int \frac{dx}{\pi a} V(x,n N a)
\sqrt{\sigma_B \sigma_V} \cos(\sqrt{2 \pi} \theta_{n,c}(x)) 
\cos(\sqrt{2 \pi} \phi_{n,s}(x))
\label{pairing}
\end{eqnarray}
and in the final Lagrangian we replace $\theta_{\rho}$ by 
$\theta_c$ and add a new term that reads:
\begin{eqnarray}
L_I[\varphi] &=& \sum_n \int dx \left\{
\frac{K_{\rho}}{4 \pi v_{\rho}}
\left[ 
\left(\partial_{\tau}(\varphi_B(x,n N a)+\varphi_V(x,n N a))\right)^2 
\right.
\right.
\nonumber
\\
&+& \left. \left. v_{\rho}^2
\left(\partial_{x}(\varphi_B(x,n N a)+\varphi_V(x,n N a))\right)^2 \right] \right.
\nonumber
\\
&+& \left. \frac{K_{\rho}}{\sqrt{2 \pi} v_{\rho}} 
\left[
\partial_{\tau}(\varphi_B(x,n N a)+\varphi_V(x,n N a)) 
\partial_{\tau}\theta_{n,c}(x) 
\right.
\right.
\nonumber
\\
&+& \left. 
\left.
v_{\rho}^2 
\partial_{x}(\varphi_B(x,n N a)+\varphi_V(x,n N a)) 
\partial_{x}\theta_{n,c}(x) \right] \right\} \, .
\label{lvarphi}
\end{eqnarray}
The physical interpretation of (\ref{thetac}) is quite interesting. Observe that the
fermion current along the stripe is 
\begin{eqnarray}
j_n(x) &=& \frac{1}{\sqrt{\pi}} \partial_x \theta_{n,\rho}(x) 
\nonumber
\\
&=& 
\frac{1}{\sqrt{\pi}} \partial_x \theta_{n,c}(x)
+\frac{1}{\pi \sqrt{2}} \partial_x \varphi_B(x,n N a)
+\frac{1}{\pi \sqrt{2}} \partial_x \varphi_V(x,n N a)
\label{currents}
\end{eqnarray}
where the first term is just the normal current, the second term is the
superfluid {\it Josephson} current and the last term is a 
current that is driven by the dislocations in the CCS. 
Moreover, observe that the bosonic field $\theta_c$ is still coupled to the phase fields
through (\ref{lvarphi}). Thus, the stripes carry along 
both superfluid and CCS distortions.

In order to proceed with the calculations we 
study the phase fluctuations around the
saddle point equations for the complete action. At the saddle point we have: $\sigma_{\alpha} 
= \sigma_{\alpha}^0$, $\varphi_{\alpha} = 0$ with $\alpha=B,V$. 
This just gives the mean-field solution given in 
Section \ref{mfield} with the difference that at the saddle point the
stripe action reads (we drop the subscript $n$):
\begin{eqnarray}
S &=& \int_0^{\beta} d\tau dx \left\{\frac{K_{\rho}}{2 v_{\rho}} \left[
(\partial_{\tau} \theta_c(x,\tau))^2 + v_{\rho}^2 (\partial_{x} \theta_c(x,\tau))^2\right]
\right.
\nonumber
\\
&+& \left. \frac{1}{2 K_s v_s} \left[
(\partial_{\tau} \phi_s(x,\tau))^2 + v_s^2 (\partial_{x} \phi_s(x,\tau))^2\right] \right\}
\nonumber
\\
&-& \int_0^{\beta} d\tau \int dx \tilde{V} 
\cos(\sqrt{2 \pi} \theta_{c}(x)) \, \, \, 
\cos(\sqrt{2 \pi} \phi_{s}(x)) \, .
\label{isolatedstripe}
\end{eqnarray}
Eq. (\ref{isolatedstripe}) 
is the interacting version of the Hamiltonian used in the previous 
Section with $\tilde{V} = \sqrt{\sigma_B^0 \sigma_V^0} V$. 
The last term in (\ref{isolatedstripe}) is the Josephson coupling
induced between different stripes by the bosons. A similar coupling
was proposed for the case where stripes cross each other \cite{prl97}.
The effect
of the pairing term can be easily seen in a renormalization group 
calculation in first order in
$V$. Repeating the RG calculation done in Section \ref{stfe} we can easily show that
the dimensionless coupling constant $v=V/E_c$ renormalizes as \cite{prl97}
\begin{eqnarray}
\frac{d v}{d \ell} = \left[2-\frac{1}{2} \left(K_s+\frac{1}{K_{\rho}}\right)\right] v(\ell) \, .
\label{vell}
\end{eqnarray}
Notice that the operator associated with $V$ is relevant  
when $K_s+K_{\rho}^{-1}<4$ and irrelevant otherwise. 
For a Hubbard model with local repulsion $U_H$ it requires 
that $U_H>1.8 t$
for the operator to become irrelevant. Thus, even if the stripes are deep inside of a CDW state
the coupling $V$ drives the system into a superconductor. 
When $V$ is large then the last
term of (\ref{isolatedstripe}) acquires a expectation value and the bosonic fields get pinned
at their minima:
\begin{eqnarray}
\theta_{c} &=& \sqrt{\frac{\pi}{2}} (n+m)
\nonumber
\\
\phi_s &=& \sqrt{\frac{\pi}{2}} (n-m)
\label{minima} 
\end{eqnarray}
where $n$ and $m$ are integers. Thus, from (\ref{currents}) one sees that the only currents
circulating along the stripes are the Josephson currents due to the superfluid and CCS fluctuations. 
Fluctuations around the minima (\ref{minima}) are massive (although a 
superfluid current still flows) and 
a spin gap opens in the spectrum. 
Indeed, a simple integration of (\ref{vell}) gives
\begin{eqnarray}
\frac{V_R}{E_R} &=& \frac{V}{E_c} \left(\frac{E_c}{E_R}\right)^{1-g}
\nonumber
\\
g &=& \frac{1}{2} \left(K_s+\frac{1}{K_{\rho}}\right) -1
\end{eqnarray}
where $V_R$ and $E_R$ are the renormalized coupling constant and bandwidth. Observe that the RG
flow stops when $V_R \approx E_R \approx m$ where 
\begin{eqnarray}
m \approx E_c \left(\frac{V}{E_c}\right)^{1/(1-g)} = V \left(\frac{V}{E_c}\right)^{g/(1-g)} 
\end{eqnarray}
gives the amplitude of the spin gap 
(when $g=1$ the operator is marginal and $m \approx E_c \exp\{-E_c/V\}$).
Since the stripe modes are massive the coupling terms
to the phase modes in (\ref{lvarphi}) are suppressed. 

The calculation can proceed in the usual way by expanding the action around the
fluctuations of the amplitudes to second order (that is, we write $\sigma_{\alpha}
=\sigma^0_{\alpha} + \delta \sigma_{\alpha}$ and integrate over 
$\delta \sigma$). The final result reads:
\begin{eqnarray}
S &=& \frac{1}{2} \int_0^{\beta}d\tau  \int d^2 r  \left\{ 
\sum_{\alpha=B,V} \left[ 
i \overline{\sigma}_{\alpha} \partial_{\tau} \varphi_{\alpha}({\bf r},\tau) 
+ \frac{\kappa_{\alpha}}{4} \left(\partial_{\tau} \varphi_{\alpha}({\bf r},\tau)\right)^2
+ \frac{\sigma_{\alpha}}{M_{\alpha}} 
\left(\nabla \varphi_{\alpha}({\bf r},\tau)\right)^2 
\right]
\right.
\nonumber
\\
&+& \left. 
\frac{K_{\rho}}{2 \pi v_{\rho} N a} 
\left[\left(\sum_{\alpha=B,V} \partial_{\tau} \varphi_{\alpha}({\bf r},\tau)
\right)^2 + v_{\rho}^2 \left(\sum_{\alpha=B,V} \partial_x \varphi_{\alpha}({\bf r},\tau)
\right)^2 
\right] 
\right\}
\label{phaseaction}
\end{eqnarray}
where the factor of $1/N$ appears because we have coarse-grained the fields in the direction perpendicular to the stripes. 

The parameters that
appear in (\ref{phaseaction}) can be obtained directly from the symmetries
of the original bosonic action (\ref{lag}). The bosons obey
periodic boundary conditions ($P({\bf r},{\beta}) = P({\bf r},0)$, 
$b({\bf r},\beta)=b({\bf r},0)$) while the fermions obey anti-periodic
boundary conditions in the imaginary time direction ($\psi({\bf r},\beta)
= - \psi({\bf r},0)$). Suppose we change the boundary conditions so that
$P({\bf r},{\beta}) = P({\bf r},0) \exp\{-i \delta \varphi\}$.
From (\ref{lag}) we can enforce the original boundary conditions
if we re-interpret the problem as a shift in the chemical potential of
the bosons from $\mu$ to $\mu' = \mu-i \delta \varphi/(2 \beta)$ 
\cite{matthew}. Because bosons and fermions are coupled, we are forced to
impose new boundary conditions for the fermions, namely:
$\psi({\bf r},\beta) = - \psi({\bf r},0) \exp\{i \delta \varphi/2\}$.
The whole action is invariant under the change
in the boundary conditions if we assume a shift in the chemical
potential. If the shift is infinitesimally small we can calculate 
the change in the free energy due to the change in the boundary
conditions:
\begin{eqnarray}
\delta F \approx \frac{\partial F}{\partial \mu} \left(-i \frac{\delta \varphi}
{2 \beta}\right) + \frac{1}{2} \frac{\partial^2 F}{\partial \mu^2} 
\left(-i \frac{\delta \varphi}{2 \beta}\right)^2
\end{eqnarray}
and since $\delta \varphi/\beta \approx \partial_{\tau} \varphi$ we
immediately find that
\begin{eqnarray}
N_s^2 a^2 \, \overline{\sigma}_B &=& 
\frac{\partial F}{\partial \mu} = \overline{N}_e
\nonumber
\\
N_s^2 a^2 \,\kappa_B &=& \frac{\partial^2 F}{\partial \mu^2}
\end{eqnarray}
where $\overline{N}_e$ is the average number of electrons in the system.
Since the number of holes in the system is $x$ we see that
$\overline{\sigma}_B = (1-x)/a^2$. Thus, we conclude that 
$\overline{\sigma}_B$ is the average planar density
of electrons. 

Notice that so far this is the
first equation where the number of holes appears explicitly.
In principle the distance between stripes, $N$, and $x$ should be
related but in our theory $N$ is an input. When doping is
increased the number of
holes in each stripe and/or the distance between stripes can change. In the
normal state the stripes are in a CDW state and there is
a gap to charge excitations because of commensurability. 
Since the holes are injected into the system
at high temperatures during the annealing of the alloy it seems reasonable
to assume that the CDW state will be charge rigid, that is, instead of
increasing the doping of individual stripes one would get more stripes but with
the same linear doping, $n_s$. If this is the case then a simple relation 
exists between $x$ and $N$, namely,
\begin{eqnarray}
x = \frac{n_{s}}{N} \, .
\label{xn}
\end{eqnarray}
The value of $n_s$ is
determined by the competition between the gain in the kinetic energy
along the stripe versus the loss of energy due to the formation of the
ADW \cite{whitescalapino,sasha}. This is not a problem we have addressed 
but DMRG calculations \cite{whitescalapino} 
and other approaches estimate that the minimization
of the energy occurs at $n_s = 1/2$ as proposed by the experiments performed
in LNSCO \cite{tranquada}.  Notice that (\ref{nm}) implies that the pseudo-gap
temperature scale, $T^*$, vanishes for $1/6<x=x_{sp}<1/4$. 
Assuming that $x_{sp} \approx 0.2$ we find that $\gamma \approx 0.5$ 
and there is a reduction of $50 \%$ of the antiferromagnetic
exchange across the stripe.

Moreover, $\kappa_B$ is charge compressibility
and the superfluid velocity is given by
\begin{eqnarray}
c_B = \frac{4 \sigma_B}{\kappa_B M_B} \, .
\end{eqnarray}
Since the fluctuations of the CCS have zero chemical
potential one concludes immediately that $\overline{\sigma}_V =0$ and
$\kappa_V$ is related to the sound velocity in the CCS that is given by
\begin{eqnarray}
c_V = \frac{4 \sigma_V}{\kappa_V M_V} \, .
\end{eqnarray}

It is convenient to rewrite the action
(\ref{phaseaction}) in Fourier space:
\begin{eqnarray}
S &=& \frac{1}{2 \beta} \sum_{{\bf k},n} \left\{
\sum_{\alpha=B,V} 
\left[-\overline{\sigma}_{\alpha} \omega_n \varphi_{\alpha}({\bf k},\omega_n)
+ \left(\kappa_{\alpha,R} \omega_n^2 +
\frac{\sigma_{\alpha}}{M_{\alpha}} \left((1+\lambda_{\alpha}) k_x^2
+  k_y^2 \right)\right)
|\varphi_{\alpha}({\bf k},\omega_n)|^2 \right]
\right.
\nonumber
\\
&+& \left. \frac{K_{\rho}}{2 \pi v_{\rho} N a}
(\omega_n^2 + v_{\rho}^2 k_x^2) \left(\varphi_B({\bf k},\omega_n) \varphi_V^*({\bf k},\omega_n) + \varphi^*_B({\bf k},\omega_n) \varphi_V({\bf k},\omega_n)
\right)
\right\}
\label{fouaction}
\end{eqnarray}
where $\omega_n = 2 \pi n/\beta$ is the Matsubara frequency and
\begin{eqnarray}
\kappa_{\alpha,R} &=& \kappa_{\alpha} + \frac{K_{\rho}}{2 \pi v_{\rho} N a}
\nonumber
\\
\lambda_{\alpha} &=& \frac{v_F M_{\alpha}}{2 \pi \sigma_{\alpha} N a}
\label{kct}
\end{eqnarray}
are the renormalized compressibility and anisotropy introduced
by the stripes (we used (\ref{vmu})). Notice that both quantities
return to unrenormalized values when the distance between stripes diverges,
that is, $N \to \infty$. As a result of the factor $\lambda_{\alpha}$ 
the superconductivity is anisotropic
and the correlation lenght, $\xi$, is direction dependent. 
It is easy to see that the ratio of the correlation lengths along the
$x$ and $y$ directions is given by
\begin{eqnarray}
\frac{\xi_{x,\alpha}}{\xi_{y,\alpha}} = \frac{1}{1+\lambda_{\alpha}} \, .
\end{eqnarray}

\begin{figure}
\epsfysize6 cm
\hspace{0cm}
\epsfbox{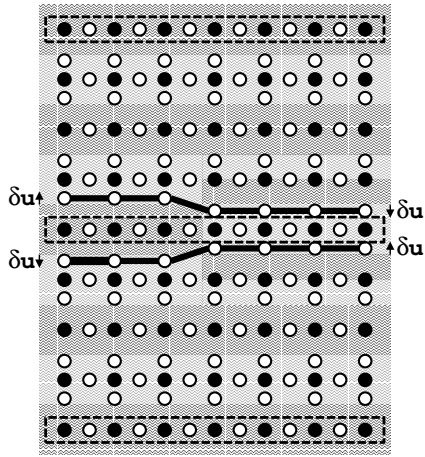}
\caption{Topological defect of the CCS.}
\label{defect}
\end{figure}

The key point of (\ref{fouaction}) is that the topological excitations of
the superfluid state (the vortices) are coupled via the
stripe compressibility to the topological excitations of the CCS (the 
dislocations). In a pictorial way consider first the ordered state
depicted on Fig.\ref{belowtc}. A defect of the ordered state 
is shown in Fig.\ref{defect} where a dislocation of the lattice 
distortion leads to a local shift of the superfluid density. In doing so
the dislocation can produce vortices in the system. The opposite situation
is also possible: vortices shift the superfluid density and drag the lattice
with them. This unusual state of affairs is the result
of the coupling of the CCS with the lattice and would not happen in ordinary
superconductors.

Although the theory described by (\ref{fouaction}) is quadratic, the nature of
the elementary topological excitations is not straightforward and we will leave
their discussion for a later publication. In this paper we 
consider a simpler problem of the effective theory for each one of the
phases. In order to do so we explore the quadratic nature of
the action (\ref{fouaction}).

\section{High temperatures}
\label{hight}

Let us consider the problem at $T_c<T<T^*$ so that the amplitude of
the order parameters are well developed but phase coherence has
not been established. It is obvious from (\ref{phaseaction}) that
this effect 
is possible due to presence of singular solutions of the field equations, that
is, due to the presence of vortex-antivortex pairs and CCS
dislocations loops. At high temperatures we can disregard the time derivatives
in (\ref{phaseaction}) since the imaginary time direction shrinks to zero
and the phase fields become independent of $\tau$. The
effective action reads:
\begin{eqnarray}
S_H = \frac{\beta}{2} \int d^2 r  \left\{ 
\sum_{\alpha=B,V}
\frac{\sigma_{\alpha}}{M_{\alpha}} 
\left(\nabla \varphi_{\alpha}({\bf r})\right)^2 
+  \frac{v_F}{2 \pi N a} 
\left(\sum_{\alpha=B,V} \partial_x \varphi_{\alpha}({\bf r})
\right)^2  
\right\} \, .
\label{phachigh}
\end{eqnarray}
Equation (\ref{phachigh}) describes two 2D XY models coupled along the 
$x$ direction. 

Let us look first at the effective field theory for $\varphi_B$ by tracing
out the $\varphi_V$ modes explicitly. This is equivalent to calculate the
renormalization of one of the modes by the gaussian fluctuations of the
other. Thus, in what follows the topological excitations of these fields
are not directly coupled to each other but only to their ``spin-waves''. This
approach is valid in the weak coupling limit of $N \gg 1$. 
It is a simple exercise to
show that the effective action becomes:
\begin{eqnarray}
S_B = \frac{\beta}{2} \sum_{{\bf k}} g_B({\bf k}) |\varphi_B({\bf k})|^2
\label{sb}
\end{eqnarray}
where 
\begin{eqnarray}
g_B({\bf k}) = {\bf k}^2 \, \, \frac{E_B E_V {\bf k}^2
+ E_C (E_B +E_V) k_x^2}{E_V {\bf k}^2 + E_C k_x^2}
\end{eqnarray}
is the effective phase propagator. We have defined 
\begin{eqnarray}
E_{\alpha} &=& \frac{\sigma_{\alpha}}{M_{\alpha}}
\nonumber
\\
E_C &=& \frac{v_F}{2 \pi N a} \, .
\end{eqnarray} 
The superfluid order parameter correlation function 
can be calculated directly from the
knowledge of (\ref{sb}). Indeed,
\begin{eqnarray}
\langle P^{\dag}({\bf r}) P(0) \rangle &=& \sigma_B \, \,  
\langle e^{i \varphi_B({\bf r})} e^{- i \varphi_B(0)} \rangle
\nonumber
\\
&=& \sigma_B \, \, \exp\left\{-\frac{1}{2} G_B({\bf r})\right\}
\end{eqnarray}
where
\begin{eqnarray}
G_B({\bf r}) = \Re\left\{\frac{1}{\beta} \sum_{{\bf k}} \frac{1-e^{i {\bf k}
\cdot {\bf r}}}{g_B({\bf k})}\right\}
\label{geba}
\end{eqnarray}
is the relevant correlation function.
The problem is simplified when we realize that it is
possible to write
\begin{eqnarray}
\frac{1}{g_B({\bf k})} = \frac{1}{E_B + E_V} \, \frac{1}{{\bf k}^2}
+ \frac{E_V/E_B}{E_V + E_B} \, \frac{1}{\left[1+E_C \left(E_B^{-1}+E_V^{-1}
\right)\right] k_x^2 + k_y^2}
\end{eqnarray}
that is the sum of the correlation function for an isotropic 2D XY model
plus the one for an anisotropic 2D XY model. The integrals can be
easily done in (\ref{geba}) and for $x \Lambda,y \Lambda \gg 1$ where
$\Lambda$ is a ultraviolet cut-off (of the order of the inverse of the
lattice spacing) we find
\begin{eqnarray}
G_B(x,y) &\approx& \frac{E_V/E_B}{2 \pi \beta (E_V +E_B) \sqrt{
1+E_C \left(E_B^{-1}+E_V^{-1}\right)}} \ln\left[\Lambda \sqrt{
\frac{x^2}{1+E_C \left(E_B^{-1}+E_V^{-1}\right)} + y^2}\right]
\nonumber
\\
&+& \frac{1}{2 \pi \beta (E_V +E_B)} \ln\left[\Lambda \sqrt{x^2 + y^2}\right]
\end{eqnarray}
which, for each direction separately, can be written as
\begin{eqnarray}
G_B(s) \approx \frac{1}{2 \pi \beta \rho_B} \ln(\Lambda |s|)
\end{eqnarray}
where $s$ can be either $x$ or $y$ and
\begin{eqnarray}
\rho_B = (E_B+E_V) \left[1+\frac{E_V/E_B}{
\sqrt{1+E_C \left(E_B^{-1}+E_V^{-1}\right)}}\right]^{-1} \, .
\label{rhob}
\end{eqnarray}
This result can be interpreted as the {\it effective superfluid stiffness} of an
isotropic 2D XY model. The error in making this approximation is equivalent
to an anisotropic change in the order parameter (that is irrelevant to
the problem of phase coherence discussed here).
An analogous calculation can be done for the phase field $\varphi_V$
and it is obvious that one has only to exchange $E_V$ by $E_B$ in the
expressions above in order to get:
\begin{eqnarray}
\rho_V = (E_B+E_V) \left[1+\frac{E_B/E_V}{
\sqrt{1+E_C \left(E_B^{-1}+E_V^{-1}\right)}}\right]^{-1}
\label{rhov}
\end{eqnarray}
that can be thought of an effective stiffness for the CCS
fluctuations. An interesting consequence of our calculations is that if 
$M_V \to \infty$ then $E_V \to 0$ and 
according to (\ref{rhov}) $\rho_V \to 0$. The
transition to the static CCS deformed phase is driven to zero
temperature. Thus, while the system becomes a superfluid-superconductor
at $T_{c,B}$, at any finite temperature dislocations of the CCS
do not allow for long-range order. 

We conclude that the transition from the ordered to the
disordered phase of the superfluid or/and the CCS is due to the 
unbinding of the topological excitations (Kosterlitz-Thouless) 
\cite{kt}: vortex-antivortex pairs
in the case of the superfluid and dislocation loops of the modulated CCS. 
The transition temperature to the ordered phase can be estimated
directly from (\ref{rhob}) and (\ref{rhov}) by \cite{kt}
\begin{eqnarray}
T_{KT,\alpha} \approx \frac{\pi}{2} \rho_{\alpha} \, 
\label{tca}
\end{eqnarray}
and the superconducting correlation length diverges as
\begin{eqnarray}
\xi_{\alpha}(T) \approx a \exp\left\{\frac{b}{\sqrt{T/T_{KT,\alpha}-1}}\right\} \, 
\label{xit}
\end{eqnarray}
where $b$ is a number of order of unit.

Although our calculation is completely 2D the transition
described here only produces quasi-long-range order. True long-range order
occurs via the coupling between planes. Thus, the real transition into
the ordered phase at finite temperatures is of the 3D-XY type. In order
to estimate the actual transition temperature, $T_c$, 
we assume that the coupling
energy per unit of lenght, $U_c$, is small compared to $T_{KT}$, 
and therefore the 2D
correlation length is well-developed when the system undergoes the phase
transition. 
The transition temperture is defined by the amount of energy required to
destroy phase coherence between two regions of size $\xi^2$ in different
planes separated by a distance $c$:
\begin{eqnarray}
T_{c,\alpha} \approx c U_c (\xi_{\alpha}(T_{c,\alpha})/a)^2 
\end{eqnarray}
that is a transcendental equation for for $T_{c,\alpha}$. Because of
the exponential dependence of $\xi$ with the temperature in (\ref{xit}) we
can solve this equation to logarithmic accuracy:
\begin{eqnarray}
T_{c,\alpha} \approx T_{KT,\alpha} \left( 1 + \frac{4 b^2}{
\ln^2(T_{KT,\alpha}/c U_c)} \right) \, .
\label{tcr}
\end{eqnarray}
This result indicates that $T_c$ depends only weakly on $c$ and it is very close
to $T_{KT}$.

\section{Zero Temperature}
\label{lowt}

When $T \to 0$ we can replace the sum over the Matsubara frequencies 
in (\ref{fouaction}) by an integral over frequency. Observe that
at small frequencies the dominant term in (\ref{fouaction}) is
the one that it is linear in the frequency. Thus, as in the case of
interacting bosons in 2D the universality of the transition is not
2D XY but 2D XY in a magnetic field \cite{matthew}. We neglect
all the quadratic terms in the frequency and retain
only the linear one. Notice that this is equivalent to assume
that $|\omega| \ll \overline{\sigma}_B/\kappa_{B,R}< 2 \pi v_{\rho} N a 
\overline{\sigma}_B/K_{\rho}$. Thus, the effect of the linear term is
to introduce a high frequency cut-off given by:
\begin{eqnarray}
\omega_c = \frac{\overline{\sigma}_B}{\kappa_{B,R}} \, .
\label{wc}
\end{eqnarray}
The equal time correlation function is:
\begin{eqnarray}
\langle P^{\dag}({\bf r},0) P(0,0) \rangle &=&
\exp\left\{-\frac{1}{2} \sum_{{\bf k},\omega} \frac{1}{g_B({\bf k})}
\left|1-e^{i {\bf k} \cdot {\bf r}}-i\overline{\sigma}_B \omega\right|^2
\right\}
\nonumber
\\
&\approx& \exp\left\{-\Re\left[\omega_c \sum_{{\bf k}} \frac{1-e^{i {\bf k} \cdot {\bf r}}}{g_B({\bf k})}
\right]
\right\}
\end{eqnarray}
where the leading order term in $\omega$ vanishes because the integral
is symmetric. Observe that this the same result of the previous section
with the temperature $T$ replaces by $2 \omega_c$! Thus, at zero
temperature the system will also have unbinding transition of topological
excitations when $N=N_c$ so that
\begin{eqnarray}
\omega_c(N_c) \approx \frac{\pi}{4} \rho_{\alpha}(N_c) \, .
\label{zteq}
\end{eqnarray}
Using(\ref{xn}), (\ref{kct}) and (\ref{wc}) 
we see that
\begin{eqnarray}
\omega_c(N) = \frac{1-\frac{1}{2 N}}{\kappa_B + \frac{K_{\rho}}{2 \pi v_{\rho} N a}}
\end{eqnarray}
is a monotonically decreasing function of $1/N$. On the other hand, from 
(\ref{rhob}):
\begin{eqnarray}
\rho_B(N) = \frac{\sigma_B(N)}{M_B} \, \frac{1+\frac{\sigma_V(N) M_B}{
\sigma_B(N)/M_V}}{1+ \frac{\frac{\sigma_V(N) M_B}{
\sigma_B(N)/M_V}}{\sqrt{1+\frac{v_F}{2 \pi N a} \left(\frac{M_B}{\sigma_B^0(N)}
+ \frac{M_V}{\sigma_V^0(N)}\right)}}}
\end{eqnarray}
that is a rather complicated expression in terms of $N$. We have seen
from (\ref{ratiovb}) that
\begin{eqnarray}
\frac{\sigma_V(N)}{\sigma_B(N)} \gg 1
\end{eqnarray}
since the electronic energies are much larger than the vibrational ones.
Assuming that this is the case we can write
\begin{eqnarray}
\rho_B(N) \approx \frac{\sigma_B(N)}{M_B} \sqrt{1+\frac{v_F}{2 \pi N a}
\frac{M_B}{\sigma_B^0(N)}}
\end{eqnarray}
that is a monotonically increasing function of $1/N$. Thus, equation
(\ref{zteq}) will have a solution at $1/N_c$ provided that 
\begin{eqnarray}
\omega_c(1/N_{sp}) > \frac{\pi}{4} \rho_{B}(1/N_{sp})
\end{eqnarray}
otherwise there would be no solution and the system would be insulating. 
Notice that $N_c<N_{sp}$ implying that at zero
temperature there there is a region $N_c<N<N_{sp}$ where superfluidity
exists even at zero temperature but without phase coherence. 
Observe that according to this
theory (even at zero temperature) there must be a region where bosons and
stripes co-exist but no long-range order is observed.

Moreover, borrowing the results from ref.\cite{matthew} we find that at
$N \approx N_c$ we must have the superconducting correlation length
diverging as
\begin{eqnarray}
\xi(N) \propto \left(\frac{1}{N}-\frac{1}{N_c}\right)^{-\nu} \, ,
\end{eqnarray}
the superfluid density behaving as
\begin{eqnarray}
\sigma_B(N) \propto \left(\frac{1}{N}-\frac{1}{N_c}\right)^{\nu z} \, ,
\end{eqnarray}
and the boson compressibility given by
\begin{eqnarray}
\kappa_B(N)\propto \left(\frac{1}{N}-\frac{1}{N_c}\right)^{\nu (2-z)} \, ,
\end{eqnarray}
where $\nu$ and $z$ are the critical exponent that at the mean-field level
are $z=1/\nu=2$. Moreover, close to $N_c$ the transition temperature scales
linearly with the superfluid density
\begin{eqnarray}
T_c(N) \propto \sigma_B(N)
\end{eqnarray} 
in complete agreement with the phenomenological Uemura relation 
observed in all superconducting cuprates \cite{uemura}. 

\section{Conclusions}
\label{conclu}

In this paper we have proposed a model of a spatially modulated collective
charge state of the cuprates where the elementary excitations change character
in real space depending on the local charge density. The problem is simplified
by assuming that there are two main kind of excitations, namely, 
Luttinger liquid degrees of freedom in the regions of high density (stripes) 
and d$_{x^2-y^2}$ bound state of fermions in the regions of lower density 
(antiferromagnetic ladders), that continously transform into each other.
We have shown that as a consequence of momentum conservation 
vibrations of the collective
state should be present in order to produce sufficient phase space for
condensation. In the presence of static or diagonal stripes superconductivity
is not possible because of the phase constraints of the former and the
d$_{x^2-y^2}$ symmetry of the boson wave-function in the latter
(in agreement with the experimental observations \cite{fujita}). 
Therefore, unlike the BCS theory 
phonons are not the basic mechanism of pairing but without vibrations
superconductivity would not be possible. 

\begin{figure}
\epsfysize10 cm
\hspace{0cm}
\epsfbox{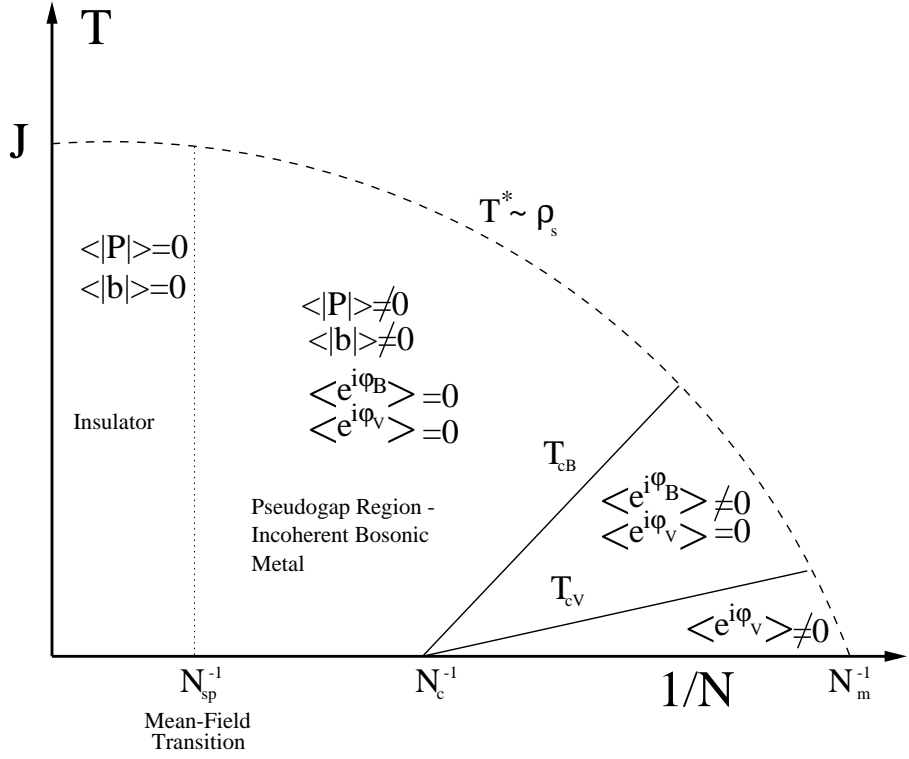}
\caption{Phase diagram of the model (the symbols are explained in the text).}
\label{finaldiag}
\end{figure}

Our main result is the phase diagram in Fig.\ref{finaldiag}. There 
$T^* \approx \rho_s(N)$ given in (\ref{rhosn}) is
the temperature scale below which d$_{x^2-y^2}$ bound states can form due
to the confining potential generated by the antiferromagnetic background.
This energy scale is set by the magnetic forces in the system and is of order
of the magnetic stiffness. Because of this constraint our
theory is only valid below the crossover line. Observe that the point where
$T^*$ vanishes at $N=N_m$ is not a quantum critical point as some theories
assume \cite{dicastro} but it sets the
scale for a crossover from 1D to 2D behavior (see (\ref{tperpn})). 
At temperatures below $T^*$
boson pairs and lattice deformations start to appear in the spectrum at
$(\pm \pi/a,0)$ and $(0,\pm \pi/a)$ but these deformations are dynamic in
nature since long range order can only be attained at low temperatures due
to phase fluctuations. Thus, the actual phase transition is driven
by topological defects of the superfluid (vortex and anti-vortex pairs) and
the CCS distortions (dislocation loops). We have shown that at finite
temperatures the transition to the ordered state is in the 3D XY universality
class but the transition temperatures for the superconductivity and static
lattice distortions can be very different because of the difference in the
stiffnesses of the phase modes. In fact, we argue that static lattice 
distortions may be observable inside of the superconducting phase at very
low temperatures. Although dynamic lattice distortions have been observed
at temperatures below $T_c$ (as given in (\ref{tcr})) 
we are not aware of observations
of static distortions at low temperatures  \cite{egami}. 
The search for such distortions
would be a good test for this theory. At zero temperature we have shown that
phase fluctuations prevent long range order to appear until the distance
between stripes reach the critical value $N_c$ (given in (\ref{zteq})) 
where phase coherence is
established. For $N_c<N<N_{sp}$ a crossover region appears in the phase
diagram where incoherent bosons co-exist with stripe fluctuations. The
phase transtion at zero temperature is in the 2D XY in a magnetic field
universality class.

\begin{figure}
\epsfysize6 cm
\hspace{0cm}
\epsfbox{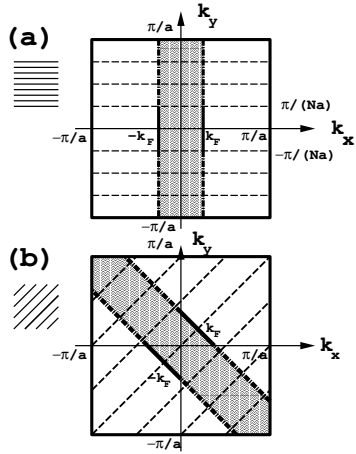}
\caption{Momentum space picture for stripes: (a) longitudinal; (b) diagonal.}
\label{bz}
\end{figure}

So far the discussion has been based on the idea of infinite 1D stripe
segments. For this geometry it is not possible to have a
true d$_{x^2-y^2}$ superconducting order parameter 
as it has been experimentally observed 
\cite{dwavexp} because the Fermi surface is flat.
For a flat Fermi surface like the one shown in Fig.\ref{bz}(a)
the superconductivity would have s-wave symmetry since the whole Fermi surface
would be gapped.
In fact this is another major problem in approaches to the cuprate problem
starting with 1D stripes, that is, how can one get a d-wave order parameter
out of a 1D problem. In that aspect our theory indeed predicts a s-wave
order parameter for infinite 1D stripes. If the small transverse tunneling
$t_{\perp}(N)$ given in (\ref{tpern}) is included, the fermion dispersion
relation is modified to (\ref{dispersion2d}) and a curvature is introduced
in the Fermi surface allowing for a very anisotropic d-wave state 
\cite{granath}. Observe,
however, that because $t_{\perp}$ is exponentially small with $N$,
changes in the Fermi surface shape will only become observable when 
$N \approx N_m$ where
the system is essentially two-dimensional. This is not sufficient to
explain the isotropic d-wave state that is observed in the underdoped
cuprates.

\begin{figure}
\epsfysize6 cm
\hspace{0cm}
\epsfbox{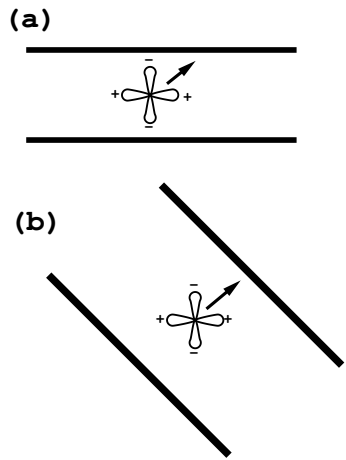}
\caption{Coupling of a d$_{x^2-y^2}$ boson wavefunction to (a) longitudinal
and (b) diagonal stripes.}
\label{cou}
\end{figure}

\begin{figure}
\epsfysize6 cm
\hspace{0cm}
\epsfbox{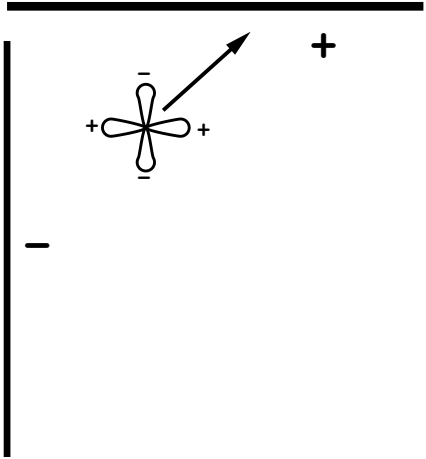}
\caption{Coupling generated by a d$_{x^2-y^2}$ boson in the presence of 
vertical and horizontal stripes.}
\label{je}
\end{figure}

\begin{figure}
\epsfysize6 cm
\hspace{0cm}
\epsfbox{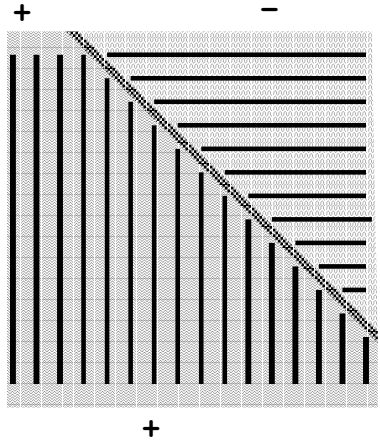}
\caption{Tunneling junction across a twin boundary between horizontal and
vertical stripes.}
\label{junction}
\end{figure}

\begin{figure}
\epsfysize6 cm
\hspace{0cm}
\epsfbox{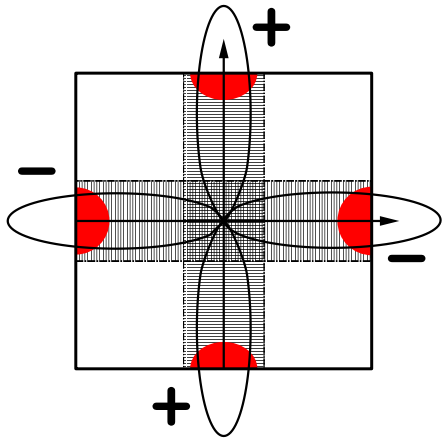}
\caption{Momentum space picture of the final d-wave state of the electrons.}
\label{dwave}
\end{figure}

\begin{figure}
\epsfysize6 cm
\hspace{0cm}
\epsfbox{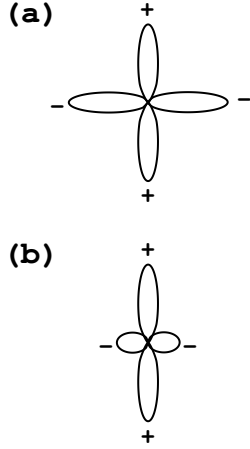}
\caption{(a) Superconducting order parameter for equal amounts of
vertical and horizontal domains; (b) superconducting order parameter
in the case where there is an excess of domains in one direction.}
\label{defdw}
\end{figure}

In reality, because of the presence of twin boundaries,
we would expect domain formation with stripes running along the crystalographic
directions (as it seems to be confirmed in neutron scattering experiments
in YBCO \cite{mook}). The existence of microscopic twins and tweeds
in the lattice structure of these systems seem to be intrinsic to
the strong lattice constraints. This situation is quite similar to what
occurs in 
some martensitic systems \cite{marten}. We can think of the 
twin boundaries as junctions between two superconductors. Since the twin 
boundary size, $L$, is very large (much larger than any of
the superconducting length scales) each mono-domain can be treated as 
a separate superconductor that is coupled to other domains through the 
twin boundary.  Because the coherence length in these systems is
of the order of the lattice spacing, the superconducting order parameter can
be continuously depressed near the twin boundary leading to a situation
similar to a superconducting-normal-superconducting junction. As we can
see in Fig.\ref{cou}(a) a d$_{x^2-y^2}$ boson couples to the positive
lobe of the wave-function to the horizontal stripes. Thus, when
phase coherence is achieved the whole Fermi surface in Fig.\ref{bz}(a)
will have the same phase sign (say, positive like in Fig.\ref{cou}(a)).
When the bosons cross the twin
boundary and find a stripe oriented along the vertical direction
the negative lobe of the wave-function couples to the vertical stripe.
This situation is shown in Fig.\ref{je}. 
Therefore adjacent domains will have opposite phases across the
twin boundary as shown in Fig.\ref{junction}.
Thus, if we superimpose the Fermi surfaces of the horizontal and vertical
stripes we find a situation like in Fig.\ref{dwave} where horizontal and
vertical Fermi surfaces will have opposite sign in their phases. 
The situation is identitical to a $\pi$-junction between two superconductors
\cite{pijunction} but the phase difference between the different
domains is zero. It is the symmetry of the boson bound state that determines
the d-wave superconductor order parameter. Notice that in the presence
of twin boundaries the electronic motion is actually 2D since stripe holes
are being transfered from vertical to horizontal stripes via the bosons.
Thus, there is true propagation along the diagonals. In the disordered phase
this propagation is not coherent and the system remains essentially 1D. In
the superconducting state, however, the coherence between the domains is
attained and quasiparticles can propagate along the diagonals as in a
ordinary superconductor. Thus, our conclusion is that in the presence
of unavoidable microscopic 
twin boundaries no quasi-particle peak is possible in the
pseudo-gap phase while in the superconducting state the quasi-particle
peak should exist. This conclusion is in agreement with the ARPES 
data \cite{shen}. A consequence of this mechanism is that the size of
the d-wave order parameter should change with the relative number of domains.
In samples where there are equal amounts of vertical and horizontal domains
the superconducting order parameter is symmetric (see Fig.\ref{defdw}(a)).
If there are 
more domains in one direction than in another
the d-wave order parameter should be assymetric with a larger lobe in
one of the directions as shown in Fig.\ref{defdw}(b). 
In fact, a s-wave component for the order parameter should also
develop since for a mono-domain system we would predict a fully
s-wave order parameter. We should also point out that c-axis coupling
can lead to further increase of the condensation energy in the system.
In fact hopping along the c-axis can help to stabilize the d-wave
order parameter \cite{paco}.

Although $T_c$ is weakly depedent on the twin boundaries in the sample,
the transport properties the critical currents and fields
will depend strongly on them. Let us consider a simple model for the
junction between a vertical and a horizontal domain. 
The free energy for the junction is written as
\begin{eqnarray}
\delta F = - \frac{I_c}{2 e} \cos(\phi)
\label{df}
\end{eqnarray}
where $I_c$ is the critical current and $\phi$ is the phase
difference between domains. 
The critical current density is
\begin{eqnarray}
J_{c} = \frac{I_c}{L \ell_c}
\label{jc}
\end{eqnarray}
where $\ell_c$ is the distance between the CuO$_2$ planes (for simplicity 
we assume that $I_{c,a} = I_{c,b} = I_c$). In the presence of an 
electromagnetic vector potential ${\bf A}$ the Josephson current 
between two different domains is simply given by
\begin{eqnarray}
I_J = 2 e \frac{\partial \delta F}{\partial \phi} = I_c \sin\left(\phi - 
\frac{2 \pi}{\Phi_0} \int_C d {\bf l} \cdot {\bf A} \right)
\label{iplane}
\end{eqnarray}
where $C$ is the line 
that links the two domains and $\Phi_0 = c/(2 e)$ is the magnetic flux 
quantum. Choosing a gauge such that $\phi = 0$ at the junction, 
assuming that the twin boundary has width $d$ and
that the variations of the order parameter across the twin boundary are 
negligible, we have from (\ref{jc}) and (\ref{iplane}):
\begin{eqnarray} 
{\bf J} =  - \frac{2 \pi J_c d}{\Phi_0} {\bf A}
\end{eqnarray}
that leads to a penetration depth, $\lambda_J$, given by
\begin{eqnarray}
\lambda_J = \sqrt{\frac{c \Phi_0}{8 \pi^2 d J_c}} \, .
\end{eqnarray}
Moreover, the variation of the phase across the twin boundary 
is $\pi/d$ but because
this variation occurs in a length scale of the correlation length induced 
by the Josephson
coupling, $\xi_J$, we expect that
\begin{eqnarray}
\xi_J \approx d \, .
\end{eqnarray}
Thus, the Ginzburg-Landau parameter, $\kappa_J$, is given by
\begin{eqnarray}
\kappa_J = \frac{\lambda_J}{\xi_J} \approx \sqrt{\frac{c \Phi_0}{8 \pi^2 d^3 J_c}}
\end{eqnarray}
leading to a critical field, $H_{c1,J}$, for the field strength that
is screened
by surface currents given by
\begin{eqnarray}
H_{c1,J} = \frac{\Phi_0}{4 \pi \lambda_J^2} \ln(\kappa_J)
\approx \frac{\pi d J_c}{c} \ln \left(\frac{c \Phi_0}{8 \pi^2 d^3 J_c}\right) \, .
\end{eqnarray}
Finally, the critical field for the penetration of one flux of quanta through the twin
boundary is of order
\begin{eqnarray}
H_{c2,J} = \frac{\Phi_0}{2 \pi \xi_J^2} \approx \frac{\Phi_0}{2 \pi d^2} \, .
\end{eqnarray}
Thus, we expect $H_{c2,J}$ to be much smaller than the upper critical field 
required for the extinction of long range order in the system.
It is clear that many macroscopic properties of the superconductor are
determined by what happens at the twin boundaries.

\begin{figure}
\epsfysize6 cm
\hspace{0cm}
\epsfbox{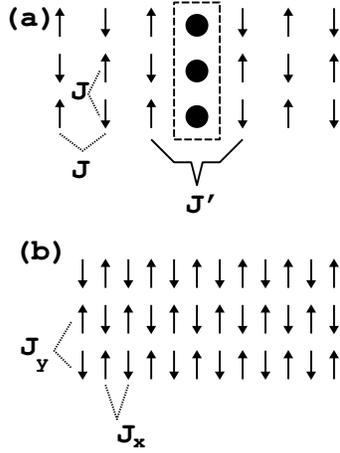}
\caption{(a) Magnetic interactions in the presence of stripes: $J$ is
the exchange between spins in the antiferromagnetic ladder; $J'$ is the
exchange across the stripe. (b) Effective magnetic model with spatially
anisotropic couplings.}
\label{anisotropy}
\end{figure}

\begin{figure}
\epsfysize6 cm
\hspace{0cm}
\epsfbox{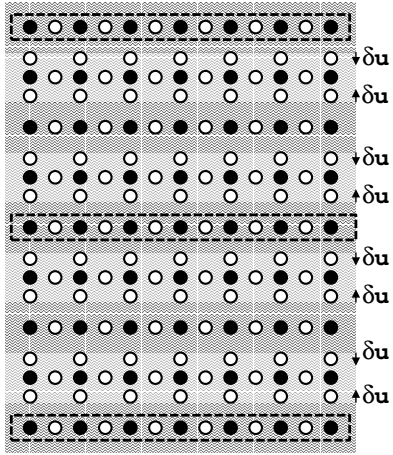}
\caption{Real space picture of the ordered phase: $\delta u$ indicates
the possible motion of the O atoms with doubling of the unit cell perpendicular
to the stripe orientation. The difference in gray levels represents the
modulation of the superfluid density.}
\label{belowtc}
\end{figure}

\begin{figure}
\epsfysize6 cm
\hspace{0cm}
\epsfbox{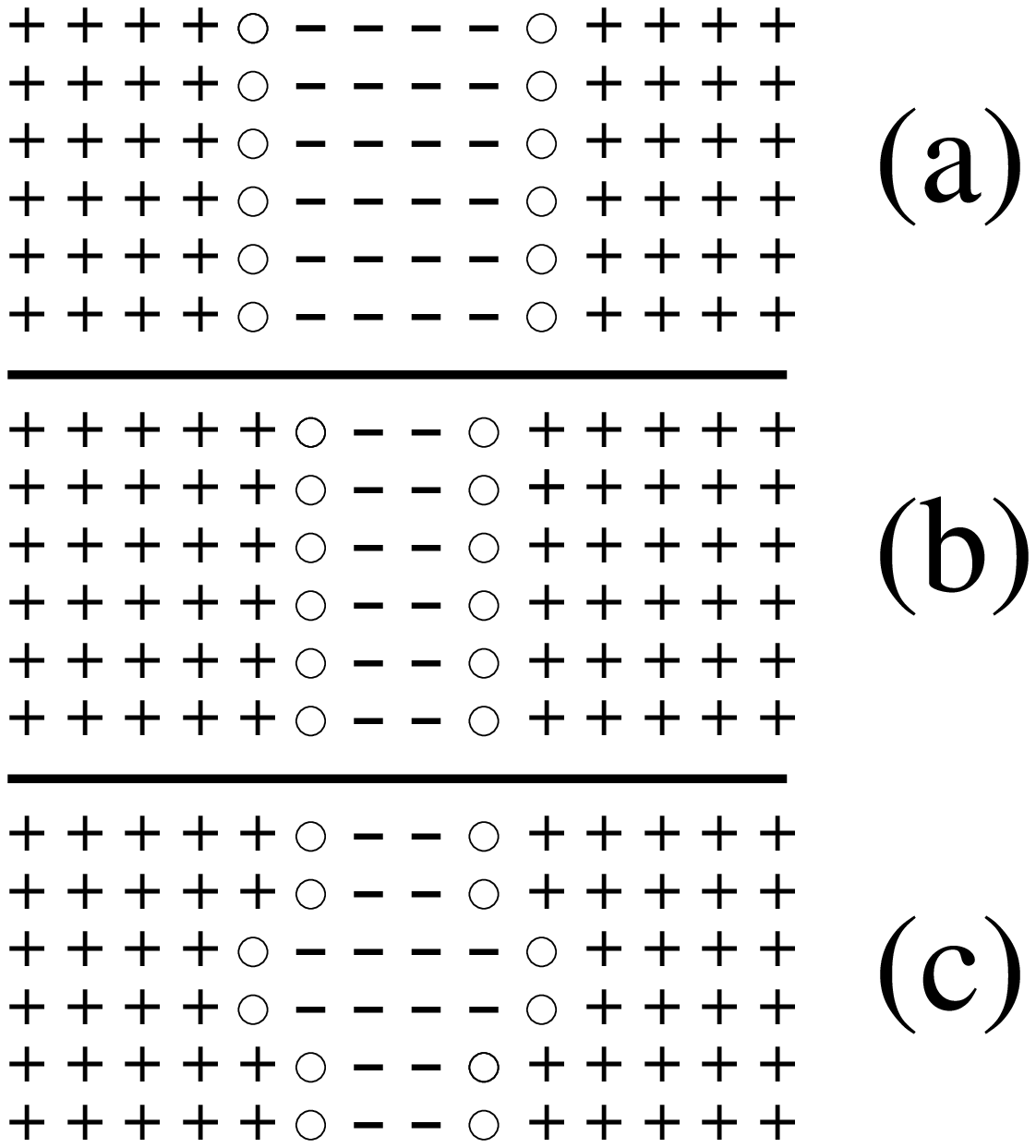}
\caption{$+$ sign indicates an up staggered magnetization, $-$ indicates
a down staggered magnetization: (a) normal phase; (b) dimerized phase;
(c) topological defect of a bond centered stripe.}
\label{comag}
\end{figure}

As we have argued the formation of the stripes is driven by the gain in
kinetic energy and therefore favors anti-phase domain walls as shown in
Fig.\ref{anisotropy}(a). Thus, the magnetic fluctuations are 
incommensurate with the lattice as it has been observed for a long time
in all the cuprates \cite{experiments}. However, we have found that
there are lattice distortions in the system that we can associate with
the O motion as shown in Fig.\ref{belowtc}. If the stripes are site
centered then the distortion does not affect the incommensurate
spin order since the Cu atoms are unaffected. However, if the the stripes
are bond centered (as some numerical \cite{alan} as well as analytical works
\cite{twor} indicate) then
the dimerization will produce {\it commensurate} magnetic response. This
can be seen in a schematic way in Fig.\ref{comag} where a dislocation
loop of the CCS produces excess of one magnetic domain (a domain with
a given staggered magnetization) over the other \cite{thankerica}. 
Thus, for bond centered
stripes the dislocation loops will produce a dynamic commensurate response. 
The presence of dynamical commensurate spin 
fluctuations have been observed in YBCO \cite{keimer} but not in LSCO.
It might well be that stripes are bond centered in YBCO and site centered
in LSCO. This would explain the difference between these two materials 
in regards to the presence of commensurate spin fluctuations at low 
temperatures. 

It is clear from our theory that the vibrations will be affected by
an isotope effect (see (\ref{du})). Moreover, because a static lattice
distortion is involved we also expect the critical temperature for
the distortions, $T_{c,V}$, to be strongly affected by changes of O isotopes
(changes of O$^{16}$ by O$^{18}$). 
There is strong evidence for the O isotope effect in LSCO and
other cuprates that support our theory \cite{alex}. Moreover, we expect
the isotope effect to be stronger at the $T=0$ transition to the
superconducting state at $N=N_c$ where the static lattice distortions
start to appear. However, the superconducting transition itself should
be very weakly dependent on the isotope effect because the binding mechanism
only involves the lattice in a indirect way (in other words, the 
stiffness of the superfluid 
is only weakly renormalized by the lattice). The experimental
data in cuprates have indeed shown an unusual isotope effect where the
critical superconducting temperature is not correlated with isotope effect
that becomes stronger at the quantum critical point associated with
superconductivity \cite{keller}.
In fact, optimally doped cuprates show very weak signs of the isotope
effect. This has been used as an argument against phonon mechanisms for
superconductivity. In our theory at $T=0$ a static lattice
distortion appears together with the superconductivity 
(see Fig.\ref{finaldiag}). It is the transition into this dimerized
stripe phase that will be strongly affected by the isotope effect, not
the superconducting one. Thus, we can explain the 
the unusual isotope effect in
cuprates as a consequence of the unavoidable coupling between lattice
and the superconducting condensate.

It is clear from the phase diagram shown in Fig.\ref{finaldiag} that
our theory does not describe the so-called overdoped region of the
cuprates. When the distance between the stripes becomes of the order
of the lattice spacing the system becomes homogeneous. The antiferromagnetic
correlation lenght, for instance, is short and the bound
states disappear from the spectrum by merging with the lower 
Hubbard band (this coincides with the vanishing
of the pseudogap energy scale). Beyond this point we believe there is
a crossover to a conventional BCS behavior with a well defined Fermi surface
and therefore to Fermi liquid behavior in the normal phase \cite{george}.
This smooth crossover is possibly the same one 
that occurs between a Bose-Einstein
system and a BCS superconductor \cite{nozieres}. 

In summary, we have presented a new model for a collective electronic
state of the cuprates where the elementary excitations change from 
place to place in real space. We show that the decaying processes among
these elementary excitations produce superconducting correlations even
when the interactions are repulsive. We show that
the d-wave nature of the order parameter is associated with the d-wave
nature of boson bound states that exist due to the magnetic confinement
in the system. We have shown that phase fluctuations are responsible
for the quantum disorder and that the phase diagram 
depends strongly on how vortices couple to dislocation loops. We have
explained various different experimental facts of the cuprates and
predicted new effects that might prove or disprove our theory.

I would like to acknowlegde fruitful conversations with A.~Bishop,
D.~Campbell,
E.~Carlson, H.~Castillo, C.~Chamon, 
A.~Chernyshev, E.~Dagotto, C.~di Castro, T.~Egami, 
E.~Fradkin, T.~H.~Geballe, N.~Hasselman, D.~MacLaughlin, A.~Moreo, 
L.~Pryadko, D.~Scalapino, Z.~X.~Shen,
C.~Morais Smith, S.~Sridhar, S.~White, and Jan Zaanen. I thank T.~H.~Geballe
for pointing out ref.\cite{bardeen}. This work was partially supported
by a CULAR-LANL grant under the auspices of the Department of Energy. 

{\bf Note}: After this paper was completed I became aware of ref.\cite{ale}
where electron-phonon coupling is discussed in the context of ARPES and
neutron scattering experiments.

$^*$ On leave from the Department of Physics, University of California,
Riverside, CA, 92521.

\newpage

\newpage

\newpage

\newpage

\newpage

\newpage

\newpage

\newpage

\newpage

\newpage

\end{document}